\documentclass[pra,onecolumn,showpacs,superscriptaddress,,preprintnumbers,amsmath,amssymb,nofootinbib]{revtex4}
\usepackage{graphicx}
\usepackage{mathtools}
\usepackage{dcolumn}
\usepackage{bm}
\usepackage{slashed}
\usepackage{ulem} 
\usepackage{color}
\usepackage[utf8x]{inputenc}
\usepackage{epstopdf}
\usepackage{empheq,fancybox}
\usepackage{relsize}

\newcommand{\bb}[1]{\mathbf{#1}}

\begin{document}
\title{Excitations and stability of weakly interacting Bose gases with 
multi-body interactions}

\author{Danny Laghi}
\affiliation{CNR-IOM DEMOCRITOS Simulation Center, Via Bonomea 265, 
I-34136 Trieste, Italy}

\author{Tommaso Macr\`i}
\affiliation{Departamento de F\'isica Te\'orica e Experimental, 
Universidade Federal do Rio Grande do Norte, 59072-970 Natal-RN,Brazil}
\affiliation{International Institute of Physics, 59078-400 Natal-RN, Brazil}

\author{Andrea Trombettoni}
\affiliation{CNR-IOM DEMOCRITOS Simulation Center, Via Bonomea 265, 
I-34136 Trieste, Italy}
\affiliation{SISSA and INFN, Sezione di Trieste, Via Bonomea 265, I-34136 
Trieste, Italy}

\begin{abstract}
We consider weakly interacting bosonic gases with local and 
non-local multi-body interactions. By using the Bogoliubov approximation, 
we first investigate contact interactions, 
studying the case in which the interparticle potential can be written 
as a sum of $N$-body $\delta$-interactions, and then considering 
general contact potentials. 
Results for the quasi-particle spectrum and the stability are presented. 
We then examine non-local interactions, focusing on two different 
cases of $3$-body non-local interactions. Our results are used   
for systems with $2$- and $3$-body $\delta$-interactions and applied 
for realistic values of the trap parameters. 
Finally, the effect of conservative $3$-body terms in dipolar systems 
and soft-core potentials (that can be simulated with Rydberg dressed atoms) 
is also studied.
\end{abstract}

\maketitle

\section{Introduction}
\label{sec:intro}

The Bogoliubov theory of weakly interacting Bose gases \cite{bogoliubov1947} 
provides an essential tool to investigate the effect of interactions 
in bosonic systems \cite{zagrebnov2001} and it plays 
a key role in the study of properties 
of Bose-Einstein condensates \cite{pethick2002,pitaevskii2016}. 
Its results for the ground-state energy 
and condensate depletion are in agreement in the weakly interacting 
limit with the findings obtained by other methods subsequently developed, 
including rigorous treatments \cite{lieb2005}. 
An important point is that also when it does not (quantitatively) 
work, it is useful to have results which are of 
guide for a qualitative understanding, 
as in the case of Helium$4$ \cite{annett2004,pitaevskii2016}, 
or to have estimates of the ground-state energy, 
as in the case of the $1D$ Lieb-Liniger model for small couplings 
\cite{lieb1961}. 
Moreover, being a self-consistent approach, where 
the number of condensate particles has 
to be self-consistently determined, it gives information 
on the issue whether there is condensation or not, 
as in low-dimensional systems \cite{pethick2002,pitaevskii2016}. 
Finally, the Bogoliubov transformation 
used to diagonalize the quadratic Hamiltonian obtained by the 
Bogoliubov approximation is used in a variety of other systems, including 
spin-wave theory of antiferromagnets \cite{manousakis1991} and 
superconductors \cite{schrieffer1964,bogoliubov1959}.

In this paper we study the generalization of the Bogoliubov 
theory to local and non-local/finite-range $3$-body and general 
multi-body interactions. Our reasons for such an investigation 
are the following: 
\begin{itemize}

\item[{\it i)}] We are firstly motivated 
by the the interest in studying the 
effects that the presence of $3$-body terms in experiments 
with ultracold atoms may induce on their equilibrium and dynamical properties, 
including the quasi-particle spectrum,  
with the goal to quantify how large are such effects. 

\item[{\it ii)}] More generally, 
when (local or non-local/finite-range) $2$-body terms
are present jointly with higher-body contributions (as $3$-body ones), 
they may compete to make the system stable or unstable and it is of interest 
to determine stability conditions and the spectrum of the quasi-particles. 
A typical 
example is given by an attractively interacting Bose gas, having $a<0$, 
which can be made stable by a repulsive $3$-body term. 

\item[{\it iii)}] Another motivation is provided by 
the dipolar gas in presence of a 
$3$-body term \cite{xi16,bisset15}. 
The effect of $3$-body interaction terms in dipolar systems 
can be very interesting. As an example, 
it was recently shown that 
for a harmonically trapped dilute dipolar condensate with a 
$3$-body short-range interaction the 
system exhibits a condensate state and a droplet state \cite{blakie2016}, 
discussing how the droplet crystal may be an excited state
arising from heating as the system crosses the phase transition.  
In the following we derive 
within the Bogoliubov theory the stability condition in presence of general 
multi-body local interactions (considering the case of 
$3$-body non-local interactions as well), then we discuss in detail some specific examples.

\item[{\it iv)}] A tool to study ultracold strongly interacting 
systems such as $1$D Bose gas \cite{lieb1961} or unitary Fermi gases 
\cite{zwerger2012} is to introduce effective local interactions. Recent 
examples are provided by the recent study of the monopole excitations 
for the $1$D Bose gas \cite{choi2015} 
using an effective Gross-Pitaevskii equation 
of the form 
\begin{equation}
i \hbar \frac{\partial \psi}{\partial t} = 
-\frac{\hbar^2}{2m} \nabla^2\psi + V_{ext} \psi + f(\rho) \psi \mbox{,}
\label{GGPE}
\end{equation}
where $V_{ext}$ is the external potential,
$\rho=\mid \psi \mid^2$ is the density and the non-linear term $f(\rho)$ 
is extracted from the solution of the Bethe ansatz integral 
equations for the (homogeneous) $1D$ Bose gas \cite{lieb1961,yang1969}. Another 
example is the study of small-amplitude Josephson oscillations 
of a $^6$Li unitary Fermi gas in a double well potential 
\cite{valtolina2015}, where experimental data were compared with 
an equation of the form \eqref{GGPE} with $\rho$ the pair density, 
$V_{ext}$ the double well one-body potential and 
$f(\rho)$ extracted from Monte Carlo numerical results \cite{valtolina2015} 
(an example of a possible parametrization of $f(\rho)$ across the BEC-BCS 
crossover is in \cite{manini2005}). It is clear that in the weakly 
interacting limit it is $f(\rho) \propto \rho$: this corresponds for the 
$1D$ Bose gas to the limit $\gamma \to 0$, where $\gamma$ is the Lieb-Liniger 
coupling constant \cite{lieb1961}, and for fermions in the BEC-BCS crossover 
to the BEC limit $a \to 0^+$, 
$a$ being the scattering length. When deviations 
from the weakly interacting limit are incorporated through a function $f$ which is 
no longer proportional to $\rho$, if the function $f$ admits a series 
expansion of the form $f=\sum_{n} c_n \rho^n$, then 
there are effective multi-body local interactions (corresponding to integer 
values $n \ge 2$). Therefore, to treat such multi-body (albeit effective) 
interaction terms one needs to study in the Bogoliubov theory such 
higher-body terms, as we do systematically below.

\end{itemize}

The paper is organized as follows. In Section \ref{sec:local} we consider 
general multi-body contact interactions for a homogeneous weakly 
interacting gas treated in the Bogoliubov approximation. We consider first the 
case of an $N$-body $\delta$-interaction and successively we consider a local 
interaction which can be expanded in series of general $N$-body terms. For 
this class of interactions we compute the spectrum, the stability condition 
and the ground-state energy (also for the $1D$ case). We finally present 
results for contact interactions that cannot be expanded in series. 
In Section \ref{sec:finite-range} 
we discuss the case of non-local interactions: after briefly reviewing the 
well-known case of a $2$-body non-local interaction, we study two different 
cases of $3$-body 
non-local interactions, and a comparison between these two cases is performed 
with a Gaussian pair-wise interaction. The results of Sections 
\ref{sec:local} and \ref{sec:finite-range} are used in Section 
\ref{sec:2_3} in the case of a model with $2$- and $3$-body 
contact interactions and the obtained findings are applied 
to possible realistic values of the trap parameters. In Section 
\ref{sec:other} we discuss some further 
realistic interaction potentials of interest for current 
experimental setups with ultracold atoms. We present results for 
a $2$-body non-local potential plus $2$- and $3$-body 
$\delta$-interactions, with applications to dipolar systems, e.g. magnetic
atoms and polar molecules, 
and soft-core potentials, that can be simulated with Rydberg dressed atoms, 
to discuss the effect of a $3$-body interaction term.
Finally we present our conclusions in Section \ref{sec:concl}, while 
more technical material is presented in the Appendices.

\section{Contact interactions}
\label{sec:local}

In this Section we consider general local $\delta$-interparticle 
potentials with multi-body interactions.

\subsection{$N$-body interaction}  
\label{sec:N-body_only}

We start by considering a model for a gas of $N_T$ bosonic particles 
interacting only via local repulsive $N$-body $\delta$-interactions. 

The general Hamiltonian for $N$-body interactions reads
\begin{align}
\hat{H} = \int \!  \mathrm{d}\bb{r} \, \hat{\Psi}^{\dagger}(\bb{r}) \, \biggl( -\frac{\hbar^2 \nabla^2}{2 m}  \biggr) \, \hat{\Psi}(\bb{r}) + \frac{1}{N!}  \int \!  \mathrm{d}\bb{r}_1 \cdots \mathrm{d}\bb{r}_N \, \hat{\Psi}^{\dagger}(\bb{r}_1) \cdots  \hat{\Psi}^{\dagger}(\bb{r}_N) \, U(\bb{r}_1,\ldots,\bb{r}_N) \, \hat{\Psi}(\bb{r}_N) \cdots \hat{\Psi}(\bb{r}_1)   \mbox{,} \label{eqn:2_bogol_1}
\end{align}
where the bosonic field operators satisfy the canonical commutation 
relations $\left[ \hat{\Psi}(\bb{r}), \hat{\Psi}(\bb{r'}) \right] 
= 0 = \left[ \hat{\Psi}^{\dagger}(\bb{r}), 
\hat{\Psi}^{\dagger}(\bb{r'}) \right]$ and 
$\left[ \hat{\Psi}(\bb{r}), \hat{\Psi}^{\dagger}(\bb{r'}) \right] 
= \delta(\bb{r} - \bb{ r'})$. 
We assume $N$-body local contact interactions having the form:
\begin{equation}
U(\bb{r}_1,\ldots,\bb{r}_N) \equiv U_N \prod_{\substack{i < j}} 
\delta(\bb{r}_i - \bb{r}_j) \mbox{,} \label{eqn:2_bogol_2}
\end{equation}
where $U_N$ is a coefficient of dimension 
$[U_N]=[E]\cdot[L]^{3N-3}$. In the usual case of 2-body 
$\delta$-interaction one has $U_2 = \frac{4 \pi \hbar^2 a}{m}$, 
where $m$ is the mass of the bosons \cite{pethick2002,pitaevskii2016}. 
With potential \eqref{eqn:2_bogol_2} the Hamiltonian reads 
\begin{equation}
\hat{H} = \int \!  \mathrm{d}\bb{r} \, \hat{\Psi}^{\dagger}(\bb{r}) \, \biggl( -\frac{\hbar^2 \nabla^2}{2 m}  \biggr) \, \hat{\Psi}(\bb{r})   + \frac{U_{N}}{N!}  \int \!  \mathrm{d}\bb{r} \, \biggl( \hat{\Psi}^{\dagger}(\bb{r}) \biggr)^{\! N} \! \! \! \cdot \! \biggl( \hat{\Psi}(\bb{r}) \biggr)^{\! N} \mbox{.} \label{eqn:2general_1_enne}
\end{equation}

Using for the field operator the expansion 
\begin{equation}
\hat{\Psi}(\bb{r}) = \sum_{\bb{p}} {\Psi}_{\bb{p}}(\bb{r}) \, \hat{a}_{\bb{p}}  \mbox{,} \label{eqn:1wave_functions}
\end{equation} 
where $\Omega = L^3$ is the volume of the system 
(chosen to be a cube of side $L$ with periodic boundary conditions) and 
${\Psi}_{\bb{p}}(\bb{r})=\frac{1}{\sqrt{\Omega}} e^{i \bb{k} \cdot \bb{r}}$, with 
$\bb{p}=\hbar \bb{k}$, the Hamiltonian assumes the form
\begin{equation}
\hat{H} = \hat{H_0} + \hat{H}_I \mbox{,}  \label{eqn:2_bogol_H}
\end{equation}
with $\hat{H_0}$ given as usual by 
\begin{equation}
\hat{H}_0 = \sum_{\bb{p}} \epsilon_{p}^0 \, \hat{a}_{\bb{p}}^{\dagger} \hat{a}_{\bb{p}} \mbox{,}
\label{kin_Ham}
\end{equation}
where $\epsilon_{p}^0 = p^2/2 m$, 
and the interaction part reading as
\begin{equation}
\hat{H}_I =  \frac{U_N}{\Omega^{N-1} N!} \! \sum_{\bigl(\sum_i \bb{p}_i = \sum_i \bb{p}_i'\bigr)} \! \! \! \! \! \hat{a}_{\bb{p}_1'}^{\dagger} \cdots \hat{a}_{\bb{p}_N'}^{\dagger} \hat{a}_{\bb{p}_N} \cdots \hat{a}_{\bb{p}_1} \mbox{,}
\end{equation}
with the definition 
\begin{equation}
\sum_{\bigl(\sum_i \bb{p}_i = \sum_i \bb{p}_i'\bigr)} \! \equiv \sum_{\substack{ \bb{p}_1, \ldots, \bb{p}_N  \\ \bb{p}_1', \ldots, \bb{p}_N' }} \! \! \delta_{\bb{p}_1 + \ldots + \bb{p}_N, \, \bb{p}_1' + \ldots + \bb{p}_N'} \mbox{.} \label{eqn:2_bogol_sum}
\end{equation}

Now, proceeding as usual, we make 
the Bogoliubov prescription \cite{bogoliubov1947,zagrebnov2001}
\begin{equation*}
\hat{a}_0 \sim  \sqrt{N_0} \mbox{.} 
\end{equation*}
As usual this implies that $N_0 \sim N_T$, where $N_0$ is the condensate 
number, or the largest eigenvalue of the one-body density matrix 
\cite{pitaevskii2016}. 
We then proceed by neglecting in $\hat{H}_I$ products of 3 or more 
$\hat{a}_{\bb{p}}^{\dagger}$ with $\bb{p} \neq 0$. 

To start with, we consider $3$-body interactions, that is, $N=3$. 
From \eqref{eqn:2_bogol_sum} it follows the total momentum conservation. 
Therefore, one has to arrange $2$ nonzero momenta between $3$ initial and 
$3$ final possible momenta. To enumerate all the possibilities, we may 
start by considering all the momenta of the creation operators equal to zero: 
\begin{center}
\begin{tabular}{l*{4}{c}r}
$\bb{p}_1'$   & $\bb{p}_2'$   & $\bb{p}_3'$   & $\bb{p}_1$  & $\bb{p}_2$   & $\bb{p}_3$  \\
\hline
 0          &0          &0          &0          & $\bb{p}$     &$-\bb{p}$  \\
  &  &  & $\bb{p}$ & 0 & $-\bb{p}$   \\
  &  &  & $\bb{p}$ & $-\bb{p}$ & 0   \\
\end{tabular}
\end{center}
i.e. the contribution to 
$\hat{H}_I$ is $N_0^2 \, \hat{a}_{\bb{p}} \hat{a}_{-\bb{p}}$ 
multiplied by $3$, that is the combinatorial multiplicity, 
$3 = \binom {3} {2} $, as illustrated in the table above. 
For a generic $N$, it is therefore straightforward to conclude 
that the coefficient $\binom {3} {2}$ has to be replaced by $\binom {N} {2}$.

Going ahead, the next possible arrangements are
\begin{center}
\begin{tabular}{l*{4}{c}r}
$\bb{p}_1'$   & $\bb{p}_2'$   & $\bb{p}_3'$   & $\bb{p}_1$  & $\bb{p}_2$   & $\bb{p}_3$  \\
\hline
 $\bb{p}$          &  0          &   0          &     $\bb{p}$     & 0     &  0  \\
  &  &  & 0 & $\bb{p}$ & 0   \\
  &  &  & 0 & 0 & $\bb{p}$   \\
0 & $\bb{p}$ & 0 & $\bb{p}$ & 0 & 0    \\
  &  &  & 0 & $\bb{p}$ & 0   \\
  &  &  & 0 & 0 & $\bb{p}$   \\
 0 & 0 & $\bb{p}$ & $\bb{p}$ & 0 & 0 \\
    &  &  & 0 & $\bb{p}$ & 0 \\
      &  &  & 0 & 0 & $\bb{p}$ \\  
\end{tabular}
\end{center}
from which one can infer that the corresponding contribution 
to $\hat{H}_I$ is $N_0^2 \, \hat{a}_{\bb{p}}^{\dagger} \hat{a}_{\bb{p}}$
with multiplicity $9$. For a general $N$ the multiplicity is $N^2$.

Finally, the remaining possibilities are when all the momenta 
of the annihilation operators are vanishing:
\begin{center}
\begin{tabular}{l*{4}{c}r}
$\bb{p}_1'$   & $\bb{p}_2'$   & $\bb{p}_3'$   & $\bb{p}_1$  & $\bb{p}_2$   & $\bb{p}_3$  \\
\hline
  $\bb{p}$         &  $-\bb{p}$       &0          &0          & 0   &  0  \\
 $\bb{p}$ & 0 & $-\bb{p}$ & &  &   \\
 0 & $\bb{p}$ & $-\bb{p}$ &  &  &    \\
\end{tabular}
\end{center}
whose contribution to the operatorial part is $N_0^2 \, \hat{a}_{\bb{p}}^{\dagger} \hat{a}_{-\bb{p}}^{\dagger}$ with multiplicity 3, as in the first case considered above; hence, for a general $N$, the multiplicity is $\binom {N} {2}$.  

Thus, the Hamiltonian \eqref{eqn:2_bogol_H} in the Bogoliubov approximation 
for a general $N$ reads
\begin{equation}
\hat{H} = \sum_{\bb{p} \neq 0} \epsilon_{p}^0 \, \hat{a}_{\bb{p}}^{\dagger} \hat{a}_{\bb{p}} 
+ \frac{U_N}{\Omega^{N-1} N!} \bigl( \sqrt{N_0} \bigr)^{2N} \! + \frac{U_N}{\Omega^{N-1} N!} \bigl( \sqrt{N_0} \bigr)^{2N-2} \sum_{\bb{p} \neq 0} \biggl\{ \binom {N} {2} \bigl( \hat{a}_{\bb{p}} \hat{a}_{-\bb{p}} + \hat{a}_{\bb{p}}^{\dagger} \hat{a}_{-\bb{p}}^{\dagger} \bigr) + N^2 \, \hat{a}_{\bb{p}}^{\dagger} \hat{a}_{\bb{p}}  \biggr\} \mbox{.}
\end{equation}
Defining the density $n \equiv \frac{N_T}{\Omega}$ and the condensate 
fraction $n_0 \equiv \frac{N_0}{\Omega}$, one gets
\begin{equation}
\hat{H} = \frac{U_N N_0^N}{\Omega^{N-1} N!} + \sum_{\bb{p} \neq 0} \biggl[ \epsilon_{p}^0 + U_N n_0^{N-1} \frac{N^2}{N !}  \biggr] \hat{a}_{\bb{p}}^{\dagger} \hat{a}_{\bb{p}} + U_N n_0^{N-1} \frac{N(N-1)}{2 N!}   \sum_{\bb{p} \neq 0} \biggl[ \hat{a}_{\bb{p}}^{\dagger} \hat{a}_{-\bb{p}}^{\dagger} + \hat{a}_{\bb{p}} \hat{a}_{-\bb{p}} \biggr] \mbox{.} \label{eqn:2_bogol_4}
\end{equation}
Introducing the total particle number operator
\begin{equation}
\hat{N} = \sum_{\bb{p}} \hat{a}_{\bb{p}}^{\dagger} \hat{a}_{\bb{p}} \mbox{,}
\end{equation}
and enforcing the total number conservation 
(or, in other terms, subtracting the chemical potential \cite{pethick2002}), 
one finally obtains 
\begin{equation}
\hat{H} = \frac{U_N n^{N-1}}{N!} N_T + \sum_{\bb{p} \neq 0}  \Biggl( \epsilon_{p}^0  +\frac{N(N \! - \! 1)}{N!} U_N n_0^{N-1} \Biggr) \hat{a}_{\bb{p}}^{\dagger} \hat{a}_{\bb{p}} + \sum_{\bb{p} \neq 0} \frac{N(N \! - \! 1)}{2 N!} U_N n_0^{N-1}  \biggl(  \hat{a}_{\bb{p}}^{\dagger} \hat{a}_{-\bb{p}}^{\dagger} + \hat{a}_{\bb{p}} \hat{a}_{-\bb{p}}  \biggr) \mbox{.}
\end{equation}
We can rewrite $\hat{H}$ as
\begin{equation}
\hat{H} = \frac{U_N N_T^{N}}{\Omega^{N-1} N!} + \! \sum_{\substack{ \bb{p} \neq 0  \\ (\bb{p}>0) }} \! \hat{H}_{\bb{p}} \mbox{,}
\end{equation}
where the sum on $\bb{p}>0$ indicates that it has to be taken 
over one half of momentum space, and 
\begin{equation}
\hat{H}_{\bb{p}} = \biggl( \! \epsilon_{p}^0 \! + \! \frac{N(N \! - \! 1)}{N!} U_N n_0^{N \! - \! 1} \biggr) \! \bigl( \hat{a}_{\bb{p}}^{\dagger} \hat{a}_{\bb{p}} + \hat{a}_{-\bb{p}}^{\dagger} \hat{a}_{-\bb{p}}  \bigr) + \frac{N(N \! - \! 1)}{N!} U_N n_0^{N-1} \bigl( \hat{a}_{\bb{p}}^{\dagger} \hat{a}_{-\bb{p}}^{\dagger} + \hat{a}_{\bb{p}} \hat{a}_{-\bb{p}}  \bigr) \mbox{.}
\end{equation}

The next step is to perform the Bogoliubov transformation
\begin{equation}\label{qp_bog} 
\begin{aligned}
\hat{a}_{\bb{p}} &= u_p \hat{\alpha}_{\bb{p}} - v_p  \hat{\alpha}_{-\bb{p}}^{\dagger} \mbox{,}\\
\hat{a}_{-\bb{p}} &= u_p \hat{\alpha}_{-\bb{p}} - v_p  \hat{\alpha}_{\bb{p}}^{\dagger} \mbox {,}
\end{aligned}
\end{equation}
where $u_p^2-v_p^2=1$ (general formulas for the coefficients $u_p$ and $v_p$ 
are given in the next Subsection). We obtain
\begin{equation}
\hat{H} = \frac{U_N}{N!} n^{N-1} N_T + \sum_{\bb{p} \neq 0 } \epsilon_{p} \, \hat{\alpha}_{\bb{p}}^{\dagger}
 \hat{\alpha}_{\bb{p}} - \frac{1}{2} \sum_{\bb{p} \neq 0} \biggl( \epsilon_{p}^0 + X^{(N)} - \epsilon_{p}  \biggr) \mbox{,} \label{eqn:2_bogol_5}
\end{equation}
where the following quantity has been introduced:
\begin{equation}
X^{(N)} \equiv \frac{N(N \! - \! 1)}{N!} U_N n_0^{N-1} \mbox{,}
\end{equation} 
so that the quasi-particle spectrum is given by 
\begin{equation}
\epsilon_{p} = \sqrt{(\epsilon_{p}^0)^2 + 2 X^{(N)} \epsilon_{p}^0} 
\mbox{.}
\label{qp_only_N}
\end{equation}
Of course, for interactions involving only $N$-body $\delta$-interactions the 
stability depends just on the sign of $X^{(N)}$: if $X^{(N)}$ is positive 
(negative), the argument in the square root of \eqref{qp_only_N} is positive 
for all $\bb{p}$ (negative for small $\bb{p}$), and the system is stable 
(unstable). When more interactions are present, then one has to impose for 
stability a suitable combination of the parameters $U_N$ to be positive, 
as discussed in the next Subsection.

\subsection{Sum of multi-body contact interactions} 
\label{sec:section1-subsection2}

We consider in this Subsection a model where 
there is a sum of $2$-body, $3$-body,...,$N$-body $\delta$-interactions 
(where $N$ is arbitrary). 
In order to generalize the formulas presented in Section 
\ref{sec:N-body_only}, we consider $N_T$ bosons 
interacting via contact repulsive interactions 
described by the Hamiltonian
\begin{equation}
\hat{H} = \int \!  \mathrm{d}\bb{r} \, \hat{\Psi}^{\dagger}(\bb{r}) \, \biggl( -\frac{\hbar^2 \nabla^2}{2 m}  \biggr) \, \hat{\Psi}(\bb{r})   + \sum_{\ell=2}^{N} \frac{U_{\ell}}{\ell!}  \int \!  \mathrm{d}\bb{r} \, \biggl( \hat{\Psi}^{\dagger}(\bb{r}) \biggr)^{\! \ell} \! \! \! \cdot \! \biggl( \hat{\Psi}(\bb{r}) \biggr)^{\! \ell} \mbox{,} \label{eqn:2general_1}
\end{equation}
where the $l$-th parameter $U_{l}$ has 
physical dimension $[E] \! \cdot \! [L]^{3(\ell-1)}$, 
whose strength refers to $l$-body interaction. 
Again using Eq. \eqref{eqn:1wave_functions}, 
the Hamiltonian \eqref{eqn:2general_1} reads 
$\hat{H} = \hat{H_0} + \hat{H}_I$ with $\hat{H_0}$ given by Eq. \eqref{kin_Ham}  
and 
\begin{equation}
\hat{H}_I = \sum_{\ell=2}^{N} \frac{U_{\ell}}{\ell!}\frac{N_0^{\ell}}{\Omega^{\ell-1}} + \sum_{\ell=2}^{N} \sum_{\bb{p} \neq 0} \frac{U_{\ell}}{\ell!}\frac{N_0^{\ell-1}}{\Omega^{\ell-1}} \biggl\{ \binom {\ell} {2} \bigl( \hat{a}_{\bb{p}} \hat{a}_{-\bb{p}} + \hat{a}_{\bb{p}}^{\dagger} \hat{a}_{-\bb{p}}^{\dagger} \bigr) + \ell^2 \, \hat{a}_{\bb{p}}^{\dagger} \hat{a}_{\bb{p}}  \biggr\} \mbox{.}   
\end{equation}

Proceeding as in Sec. (\ref{sec:N-body_only}), the Hamiltonian takes the form
\begin{equation}
\hat{H} = \sum_{\ell=2}^{N}  \frac{U_{\ell}}{\ell!}\frac{N_T^{\ell}}{\Omega^{\ell-1}} +  
\sum_{\bb{p} \neq 0}  \epsilon_{p}^0  \hat{a}_{\bb{p}}^{\dagger} \hat{a}_{\bb{p}} + 
\sum_{\ell=2}^{N} \sum_{\bb{p} \neq 0} \Biggr\{ \frac{\ell(\ell-1)}{\ell!} U_{\ell} \, n_0^{\ell-1} \hat{a}_{\bb{p}}^{\dagger} \hat{a}_{\bb{p}} + \frac{\ell(\ell-1)}{2 {\ell!}} U_{\ell} \, n_0^{\ell-1}  \Biggl(  \hat{a}_{\bb{p}}^{\dagger} \hat{a}_{-\bb{p}}^{\dagger} + \hat{a}_{\bb{p}} \hat{a}_{-\bb{p}}  \Biggr) \Biggr\} \mbox{.}
\end{equation}
It follows
\begin{equation}
\hat{H} = \sum_{\ell=2}^{N}  \frac{U_{\ell}}{\ell!}\frac{N_T^{\ell}}{\Omega^{\ell-1}} + \sum_{\substack{ \bb{p} \neq 0  \\ (\bb{p}>0) }}  \biggl\{ \! \biggl( \epsilon_{p}^0  + X \biggr) \biggl( \hat{a}_{\bb{p}}^{\dagger} \hat{a}_{\bb{p}} + \hat{a}_{-\bb{p}}^{\dagger} \hat{a}_{-\bb{p}} \biggr) + X \biggl(  \hat{a}_{\bb{p}}^{\dagger} \hat{a}_{-\bb{p}}^{\dagger} + \hat{a}_{\bb{p}} \hat{a}_{-\bb{p}}  \biggr) \! \biggr\} \mbox{,}
\end{equation}
where
\begin{equation}
X = \sum_{\ell=2}^{N} \frac{\ell(\ell-1)}{\ell!} U_{\ell} \, n_0^{\ell-1} = \sum_{\ell=2}^{N} X^{(\ell)} 
\mbox{.} \label{eqn:definition_X_N}
\end{equation} 
The quasi-particles are introduced according to Eqs. \eqref{qp_bog}, and it has to be 
$u_p^2 - v_p^2 = 1$ in order to guarantee the commutation relations 
$\bigl[ \hat{a}_{\pm \bb{p}}, \hat{a}_{\pm \bb{p'}}^{\dagger}  \bigr] = \delta_{\bb{p} \bb{p'}}$. With the 
parametrization $u_p=\cosh{t}$, $v_p=\sinh{t}$, one gets
\begin{equation}
\tanh{2t} = \frac{X}{\epsilon_{p}^0 + X} \mbox{,}
\end{equation}
from which follows
\begin{equation}
u_p^2 = \frac{1}{2} \biggl( \frac{\xi_p}{\epsilon_p} + 1 \biggr) \mbox{,} \qquad v_p^2 = \frac{1}{2} \biggl( \frac{\xi_p}{\epsilon_p} - 1 \biggr) \mbox{,}
\end{equation}
where $\xi_p = \epsilon_p^0 + X$ and  
\begin{equation}
\epsilon_{p} = \sqrt{(\epsilon_p^0)^2 + 2 X \epsilon_p^0} \label{eqn:2excitation_spectrum}\mbox{.}
\end{equation}
The diagonalization yields
\begin{equation}
\hat{H} = \sum_{\ell=2}^{N}  \frac{U_{\ell}}{\ell!}\frac{N_T^{\ell}}{\Omega^{\ell-1}} + \sum_{\bb{p} \neq 0} \epsilon_{p} \, \hat{\alpha}_{\bb{p}}^{\dagger} \hat{\alpha}_{\bb{p}} - \frac{1}{2} \sum_{\bb{p} \neq 0} \Bigl( \epsilon_{p}^0 + X - \epsilon_{p}  \Bigr) \mbox{.} \label{eqn:2general_diagonalized}
\end{equation}
The excitation spectrum \eqref{eqn:2excitation_spectrum} for small $p$ 
gives 
$\epsilon_p = \mathfrak{s} p$, with the sound velocity $\mathfrak{s}$ given by
\begin{equation}
\mathfrak{s}^2 = \frac{X}{m} \mbox{.}
\end{equation}

The stability condition can be deduced from the sign of $X$, stability requiring $X>0$.

\subsection{Depletion at $T=0$ and ground-state energy}

The density of particles in the excited states is
\begin{equation}
n_{ex} = \frac{1}{V} \sum_{\bb{p} \neq 0} v_p^2 = \frac{1}{3 \pi^2} \biggl( \frac{\sqrt{m \, X}}{\hbar} \biggr)^{\! 3} \mbox{,} \label{eqn:excited_state}
\end{equation}
so that the depletion fraction can be written as
\begin{equation}
1-\frac{n_0}{n}=\frac{n_{ex}}{n} = \frac{1}{3 \pi^2 n} \biggl( \frac{m \, 
\mathfrak{s}}{\hbar} \biggr)^{\! 3} \mbox{.}
\end{equation}
The previous expression is the usual one from the $2$-body contact interaction with the substitution 
$U_2 n_0 \to X$: notice however that if one wants to use it to determine self-consistently $n_0/n$ via the 
relation $n_0/n=1-n_{ex}/n$, one has to take into account the dependence of the coefficient 
$X$ (entering $\mathfrak{s}$) on the condensate density $n_0$ according to Eq. \eqref{eqn:definition_X_N}.

To compute the ground-state energy $E_0$ in $3D$ one has to regularize 
the large-$p$ divergence \cite{pethick2002,pitaevskii2016}. 
The correct way to write the ground-state energy is
\begin{equation}
E_0 = \sum_{\ell=2}^{N}  \frac{U_{\ell}}{\ell!} \frac{N_T^2}{\Omega}   \biggl( \frac{N_T}{\Omega} \biggr)^{\! \ell-2} \! \! \! \! \! \! - \frac{1}{2} \, \sum_{\bb{p}} \Biggl( \epsilon_{p}^0 + X - \epsilon_{p} - \frac{X^2}{2 \epsilon_p^0}  \Biggr) \mbox{,} 
\end{equation} 
where we used $n_0 \approx n$ 
and the sum over $\bb{p}$ is up to the cut-off scale 
\cite{pethick2002}. Finally we get:
\begin{equation}
\frac{E_0}{\Omega} = \sum_{\ell=2}^{N}  \frac{U_{\ell}}{\ell!} n^{\ell} + \frac{8}{15 \pi^2} \biggl( \frac{m \, \mathfrak{s}}{\hbar} \biggr)^{\! 3} \mbox{.}        \label{eqn:3_ground-state}
\end{equation}

\subsection{$1D$ case}

The computation presented in the previous Subsections applies as well to the 
one-dimensional Hamiltonian
\begin{equation}
\hat{H} = \int \!  \mathrm{d}x \, \hat{\Psi}^{\dagger}(x) \, \biggl( \! -\frac{\hbar^2}{2 m}\frac{\partial^2}{\partial x^2}  \biggr) \, \hat{\Psi}(x) \, + \sum_{\ell=2}^{N} \frac{U_{\ell}}{\ell!} \! \int \!  \mathrm{d}x \, \biggl( \hat{\Psi}^{\dagger}(x) \biggr)^{\! \ell} \! \! \cdot \! \biggl( \hat{\Psi}(x) \biggr)^{\! \ell} \mbox{,} 
\end{equation}
(of course no finite condensate fraction is obtained in $1D$). 
Denoting in the $1D$ case the length of the system by $L$ and the particle density by 
$\rho \equiv \frac{N_T}{L}$, one gets the ground-state energy 
\begin{equation}
E_0 \! = \sum_{\ell=2}^N \! \frac{U_{\ell}}{\ell!} \frac{N_T^2}{L} \biggl( \frac{N_T}{L} \biggr)^{\! \ell-2} \! \! \! \! \! \! \!
  - \, \frac{1}{2} \sum_{p} \Bigl( \epsilon_{p}^0 + X - \epsilon_{p}  \Bigr) \mbox{,}
\end{equation}
where $\epsilon_p^0 = p^2/2 m$, $\epsilon_p = \sqrt{\bigl( \epsilon_p^0 \bigr)^2 + 2 X \epsilon_p^0} \,$ and 
\begin{equation}
X = \sum_{\ell=2}^N \frac{\ell(\ell-1)}{\ell!} U_{\ell} \, \rho^{\ell-1} \mbox{.}
\end{equation}
In this way the ground-state energy becomes
\begin{equation}
E_0 = \sum_{\ell=2}^N \frac{U_{\ell}}{\ell!}\rho^{\ell-1} N_T - \frac{L}{2 \pi \hbar} \int_0^{\infty} \! \! \! \! \! \mathrm{d}p \biggl[ \frac{p^2}{2m} + X - \sqrt{\frac{p^2}{2 m} + 2 X \epsilon_p^0 }\biggr]\mbox{.} \label{eqn:GS_1D}
\end{equation} 
Defining
\begin{equation}
\gamma_{\ell} = \frac{U_{\ell}}{\ell! \, \rho} \mbox{,}
\end{equation}
after calculating the integral in Eq. \eqref{eqn:GS_1D}, which converges to a finite nonzero value, 
the ground-state energy per particle in the Bogoliubov approximation is found to be
\begin{equation}
\frac{E_0}{N} = \sum_{\ell=2}^N \gamma_{\ell} \, \rho^{\ell} - \frac{2}{3 \pi}\frac{\sqrt{m}}{\hbar \, \rho} \, 
X^{3/2} \mbox{.} 
\end{equation}
This result is the generalization 
up to $N$-body contact interactions, of the Bogoliubov 
result obtained in \cite{lieb1961} for only $2$-body 
repulsive $\delta$-interactions in $1D$, which for 
small values of $\gamma_2$ is in agreement with the exact result 
\cite{lieb1961}.

\subsection{General multi-body contact interactions} 
\label{sec:general}

In this Subsection we briefly discuss two further generalizations 
of the Bogoliubov theory to two local interaction potentials.

In Section \ref{sec:N-body_only} we considered a Hamiltonian of the form
\begin{equation}
\hat{H} = \int \!  \mathrm{d}\bb{r} \, \hat{\Psi}^{\dagger}(\bb{r}) \, \biggl( -\frac{\hbar^2 \nabla^2}{2 m}  \biggr) \, \hat{\Psi}(\bb{r})   + \frac{U_{\ell}}{\ell!}  \int \!  \mathrm{d}\bb{r} \, \biggl( \hat{\Psi}^{\dagger}(\bb{r}) \biggr)^{\! \ell} \! \! \! \cdot \! \biggl( \hat{\Psi}(\bb{r}) \biggr)^{\! \ell} \mbox{,} \label{eqn:2general_1_bis}
\end{equation}
with $\ell=N$ integer and larger or equal than $2$. If $\ell$ is a real number 
(with $\ell>1$) and using in Eq. \eqref{eqn:2general_1_bis} $\Gamma(l+1)$ instead of $\ell!$ 
($\Gamma$ being the Gamma function), 
then one can show that in the Bogoliubov 
approximation the following form for the Hamiltonian still holds:
\begin{equation}
\hat{H} = \frac{U_{\ell}}{\ell!}\frac{N_T^{\ell}}{\Omega^{\ell-1}} + \! \sum_{\substack{ \bb{p} \neq 0  \\ (\bb{p}>0) }}  \biggl\{ \! \biggl( \epsilon_{p}^0  + 
X^{(\ell)} \biggr) \biggl( \hat{a}_{\bb{p}}^{\dagger} \hat{a}_{\bb{p}} + \hat{a}_{-\bb{p}}^{\dagger} \hat{a}_{-\bb{p}} \biggr) + X^{(\ell)} \biggl(  \hat{a}_{\bb{p}}^{\dagger} \hat{a}_{-\bb{p}}^{\dagger} + \hat{a}_{\bb{p}} \hat{a}_{-\bb{p}}  \biggr) \! \biggr\} \mbox{,}
\end{equation}
where
\begin{equation}
X^{(\ell)} = \frac{\ell(\ell-1)}{\Gamma(\ell+1)} U_{\ell} \, n_0^{\ell-1} \mbox{.} 
\label{eqn:definition_X_N_real}
\end{equation} 
The results of Section \ref{sec:section1-subsection2} still hold 
with $X^{(\ell)}$ instead of $X$, i.e. the quasi-particle spectrum is 
given by $\epsilon_{p} = \sqrt{(\epsilon_p^0)^2 + 2 X^{(\ell)} \epsilon_p^0}$.

Finally, we may consider a general contact Hamiltonian of the form
\begin{equation}
\hat{H} = \int \!  \mathrm{d}\bb{r} \, \hat{\Psi}^{\dagger}(\bb{r}) \, \biggl( -\frac{\hbar^2 \nabla^2}{2 m}  \biggr) \, \hat{\Psi}(\bb{r})   + 
\int \!  \mathrm{d}\bb{r} \, \colon  \!
{\cal F} \!\left( \hat{\rho} \right) \! \colon \mbox{,} \label{eqn:2general_1_ter}
\end{equation}
with $\hat{\rho}=\hat{\rho}(\bb{r})=\hat{\Psi}^{\dagger}(\bb{r}) \hat{\Psi}(\bb{r})$ and 
${\cal F}$ a function of the density operator, 
with the normal ordering 
to be taken in the second term of the right-hand side 
of \eqref{eqn:2general_1_ter}. Details are given in Appendix \ref{sec:spectrum_general_F}. One gets
\begin{equation} \label{eqn:H_general_X}
\hat{H} = {\cal E} + \! \sum_{\substack{ \bb{p} \neq 0  \\ (\bb{p}>0) }}  \biggl\{ \! \biggl( \epsilon_{p}^0  + 
X \biggr) \biggl( \hat{a}_{\bb{p}}^{\dagger} \hat{a}_{\bb{p}} + \hat{a}_{-\bb{p}}^{\dagger} \hat{a}_{-\bb{p}} \biggr) + X \biggl(  \hat{a}_{\bb{p}}^{\dagger} \hat{a}_{-\bb{p}}^{\dagger} + \hat{a}_{\bb{p}} \hat{a}_{-\bb{p}}  \biggr) \! \biggr\} \mbox{,}
\end{equation}
where ${\cal E}/\Omega={\cal F}\! \left({n}\right)$ and the parameter $X$ 
entering the quasi-particle spectrum 
$\epsilon_{p} = \sqrt{(\epsilon_p^0)^2 + 2 X \epsilon_p^0}$ given by
\begin{equation}
X = n_0 \cdot \frac{\partial^2 {\cal F} \! \left( n \right)}
{\partial n^2}\Bigg\vert_{n=n_0} \mbox{.} 
\label{eqn:definition_X_N_general}
\end{equation} 

\section{Non-local interactions}
\label{sec:finite-range}

In the previous Section 
we considered general multi-body contact interactions, showing that 
the well-known results for the $2$-body $\delta$-interactions in the homogeneous case 
are generalized in the Bogoliubov approximation by the substitution $U_2 n_0 \to X$, where 
$X$ is given by Eq. \eqref{eqn:definition_X_N} 
[or, according to the considered case, by \eqref{eqn:definition_X_N_real} 
or \eqref{eqn:definition_X_N_general}]. 
In principle, one should determine self-consistently 
$n_0/n$ from Eq. \eqref{eqn:excited_state}, but since the Bogoliubov approximation 
works when $n_0 \approx n$, 
then one can make the substitution $n_0 \to n$ in $X$, resulting 
for a sum of $N$-body contact interactions 
in the substitution $U_2 n \to \sum_{\ell} U_{\ell} \ell (\ell-1) n^{\ell-1}/
\ell!$. The case of higher-body non-local interactions is instead different and the final 
result (e.g., the quasi-particle spectrum) depends on the specific form of the interactions. 
In the following we explicitly show this for two different cases of $3$-body 
non-local interactions, 
cases that we treat after briefly recalling the corresponding well-known results for $2$-body non-local interactions.

\subsection{$2$-body non-local potential}  \label{sec:2-body_finite}

We start considering a Hamiltonian for a gas of $N_T$ bosons 
in a region of volume $\Omega$ and interacting via a $2$-body 
local repulsive, non-local potential $V_2(\bb{r})$:
\begin{equation}
\hat{H} = \int \!  \mathrm{d}\bb{r} \, \hat{\Psi}^{\dagger}(\bb{r}) \, \biggl( \! -\frac{\hbar^2 \nabla^2}{2 m}  \biggr) \, \hat{\Psi}(\bb{r})  +  \frac{1}{2!} \int \!  \mathrm{d}\bb{r}_1 \mathrm{d}\bb{r}_2 \, \hat{\Psi}^{\dagger}(\bb{r}_1)  \hat{\Psi}^{\dagger}(\bb{r}_2) \, V_2(\bb{r}_1 - \bb{r}_2) \, \hat{\Psi}(\bb{r}_2) \hat{\Psi}(\bb{r}_1)   \mbox{,} \label{eqn:4_short_1}
\end{equation}
with $V_2(\bb{r})=V_2(-\bb{r})$. In momentum space, the Hamiltonian \eqref{eqn:4_short_1} reads 
$\hat{H} = \hat{H_0} + \hat{H}_I$ 
with $\hat{H_0}$ given by Eq. \eqref{kin_Ham} and 
\begin{equation}
\hat{H}_I = \sum_{\substack{\bb{p}_1 \bb{p}_2 \\ \bb{p}_1' \bb{p}_2'}}  \frac{1}{2 \Omega^2} \int \!  \mathrm{d}\bb{r}_1 \mathrm{d}\bb{r}_2 \, V_2(\bb{r}_1 - \bb{r}_2)  \, \hat{a}^{\dagger}_{\bb{p}_1'} \hat{a}^{\dagger}_{\bb{p}_2'} \hat{a}_{\bb{p}_1} \hat{a}_{\bb{p}_2} \, e^{-\frac{i}{\hbar} (\bb{p}_1 \!- \bb{p}_1') \cdot \bb{r}_1} \, e^{-\frac{i}{\hbar} (\bb{p}_2 - \bb{p}_2') \cdot \bb{r}_2}  \mbox{.}
\end{equation}
With the change of variables $\bb{r} = \bb{r}_1 - \bb{r}_2$, $\bb{R} = \frac{\bb{r}_1 + \bb{r}_2}{2}$ 
and using the Bogoliubov approximation, the Hamiltonian \eqref{eqn:4_short_1} becomes
\begin{equation}
\hat{H} = \frac{n V_0}{2}N_T + \! \sum_{\substack{ \bb{p} \neq 0  \\ (\bb{p}>0) }} \hat{H}_{\bb{p}} \mbox{,}
\end{equation}
where
\begin{equation}
\hat{H}_{\bb{p}} = \Bigl(   \epsilon_p^0 + n_0 V_{\bb{p}}  \Bigr) \Bigl( \hat{a}_{\bb{p}}^{\dagger} \hat{a}_{\bb{p}} + \hat{a}_{-\bb{p}}^{\dagger} \hat{a}_{-\bb{p}} \Bigr) + n_0 V_{\bb{p}} \Bigl( \hat{a}_{\bb{p}}^{\dagger} \hat{a}_{-\bb{p}}^{\dagger} + \hat{a}_{\bb{p}} \hat{a}_{-\bb{p}} \Bigr) \mbox{,}
\label{H_p_fr}
\end{equation}
having introduced the Fourier transform $V_{\bb{p}}$:
\begin{equation} \label{eqn:F.T. definition}
V_{\bb{p}} = \int \!  \mathrm{d}\bb{r} \, V_2(\bb{r}) \, e^{-\frac{i}{\hbar} \bb{p} \cdot \bb{r}} \mbox{.}
\end{equation}

The Hamiltonian \eqref{H_p_fr} is readily diagonalized obtaining
\begin{equation}
\hat{H} = \frac{n V_0}{2}N_T + \sum_{\bb{p} \neq 0} \epsilon_{p} \, \hat{\alpha}_{\bb{p}}^{\dagger} \hat{\alpha}_{\bb{p}} - \frac{1}{2} \sum_{\bb{p} \neq 0} \biggl( \epsilon_{p}^0 + n_0 V_{\bb{p}} - \epsilon_{\bb{p}}  \biggr) \mbox{,} \label{eqn:4general_diagonalized}
\end{equation}
where the excitation spectrum is now
\begin{equation}
\epsilon_p = \sqrt{(\epsilon_p^0)^2 + 2 n_0 V_{\bb{p}} \epsilon_p^0} \mbox{.} \label{eqn:spectrum_2body}
\end{equation}
To conclude this Section we mention that 
a detailed discussion of the non-local interactions 
in Bogoliubov approximation is reported in the recent paper \cite{yukalov16}, 
while a study of the two-body problem with arbitrary finite-range interactions 
on a lattice is in \cite{valiente2010}.

\subsection{$3$-body non-local potentials}

According to Eq. \eqref{eqn:2_bogol_1}, 
for $3$-body interactions the Hamiltonian reads in general $\hat{H} = \hat{H}_0 + \hat{H}_I$,  
where
\begin{equation}
\hat{H}_0 = \int \!  \mathrm{d}\bb{r} \, \hat{\Psi}^{\dagger}(\bb{r}) \, \biggl( \! -\frac{\hbar^2 \nabla^2}{2 m}  \biggr) \, \hat{\Psi}(\bb{r}) \mbox{,} 
\end{equation}
and 
\begin{equation}
\hat{H}_I = \frac{1}{3!}  \int \!  \mathrm{d}\bb{r}_1 \, \mathrm{d}\bb{r}_2  \, \mathrm{d}\bb{r}_3  \, \hat{\Psi}^{\dagger}(\bb{r}_1) \hat{\Psi}^{\dagger}(\bb{r}_2) \hat{\Psi}^{\dagger}(\bb{r}_3) \, U(\bb{r}_1,\bb{r}_2,\bb{r}_3) \, \hat{\Psi}(\bb{r}_3) \hat{\Psi}(\bb{r}_2) \hat{\Psi}(\bb{r}_1) \mbox{.} 
\end{equation}
In the following we consider two different kinds of non-local 
potentials and derive their excitation spectrum.

\subsubsection{Potential as a sum of terms with $2$ factors} \label{sub_sub_1}

We consider a potential of the form
\begin{equation}
U(\bb{r}_1,\bb{r}_2,\bb{r}_3) \equiv \frac{1}{3} \, \sum_{i=1}^{3}  \prod_{\substack{j=1 \\ j \neq i}}^{3} V(\bb{r}_i -\bb{r}_j) \mbox{,} \label{eqn:potential_2factors}
\end{equation}
where the dimension of the $V$ entering Eq. \eqref{eqn:potential_2factors} is 
$[V]=[E]^{1/2}$ [with $V(\bb{r})=V(-\bb{r})$]. 
The Hamiltonian in the Bogoliubov approximation reads
\begin{equation}
\hat{H} = \frac{N_0}{6} n_0^2 V_{0}^{2} + \sum_{\bb{p} \neq 0} \biggl\{ \biggl[ \, \epsilon_p^0  +  \frac{n_0^2}{6} \Bigl( 3 V_0^2 + 4 V_0 V_{\bb{p}} + 2V_{\bb{p}}^2 \Bigr) \biggr] \hat{a}_{\bb{p}}^{\dagger} \hat{a}_{\bb{p}} +  \frac{n_0^2}{6} \biggl[ 2V_0 V_{\bb{p}} + V_{\bb{p}}^2 \biggr] \! \Bigl( \hat{a}_{\bb{p}}^{\dagger} \hat{a}_{-\bb{p}}^{\dagger} + \hat{a}_{\bb{p}} \hat{a}_{-\bb{p}} \Bigr) \! \!\biggr\} \mbox{,}
\end{equation}
where we used the convention \eqref{eqn:F.T. definition} for the Fourier transform (and we denote 
$V_{\bb{p}=0}$ by $V_0$). After some further manipulations, the final result is
\begin{equation}
\hat{H} = \frac{n^2 V_0^2}{6} N_T + \sum_{\substack{ \bb{p} \neq 0  \\ (\bb{p}>0) }} H_{\bb{p}} \mbox{,} \label{eqn:6H_total}
\end{equation}
where
\begin{equation}
H_{\bb{p}} = \Bigl( \epsilon_k^0 + X^{(3)}_{\bb{p}} \Bigr) \Bigl( \hat{a}_{\bb{p}}^{\dagger} \hat{a}_{\bb{p}} + \hat{a}_{-\bb{p}}^{\dagger} \hat{a}_{-\bb{p}} \Bigr) + X^{(3)}_{\bb{p}} \Bigl( \hat{a}_{\bb{p}}^{\dagger} \hat{a}_{-\bb{p}}^{\dagger} + \hat{a}_{\bb{p}} \hat{a}_{-\bb{p}} \Bigr) \mbox{,}
\end{equation}
with
\begin{equation}
X^{(3)}_{\bb{p}} \equiv \frac{n_0^2}{6} \bigl( 4 V_0 V_{\bb{p}} + 2 V_{\bb{p}}^2   \bigr) \mbox{.}
\end{equation}
After diagonalizing \eqref{eqn:6H_total}, the quasi-particle energy spectrum $\epsilon_{\bb{p}}$ is seen to be
\begin{equation}
\epsilon_{\bb{p}} = \sqrt{\	\bigl(\epsilon_p^0\bigr)^2 + 2 \, X^{(3)}_{\bb{p}} \, \epsilon_p^0} = \sqrt{\bigl(\epsilon_p^0\bigr)^2 + \frac{2}{3} n_0^2 \Bigl( 2 V_0 V_{\bb{p}} + V_{\bb{p}}^2 \Bigr) \, \epsilon_p^0} \mbox{.}
\end{equation}

\subsubsection{Potential as a product of $3$ factors} \label{sec:3-body_3-factors}

The potential \eqref{eqn:potential_2factors} is a sum of three terms, each of them given by the possible pairs which can be formed between the particles. One can also consider a potential which is the 
product of the three pair interactions:
\begin{equation}
U(\bb{r}_1,\bb{r}_2,\bb{r}_3) = V(\bb{r}_1 -\bb{r}_2)V(\bb{r}_2 - \bb{r}_3)V(\bb{r}_3 - \bb{r}_1) \mbox{,}
\label{three_fact}
\end{equation}
where the single factor has now dimension $[V]=[E]^{1/3}$, again 
$V(\bb{r}) = V(-\bb{r})$. 
The Fourier transform $V_{\bb{p}}$, defined in Eq. \eqref{eqn:F.T. definition}, has dimension 
$[V_{\bb{p}}]=[E]^{1/3} \cdot [L]^3$. 

The interaction Hamiltonian is written as
\begin{equation}
\hat{H}_I =  \frac{1}{6 \, \Omega^3} \sum_{\bb{p}_1', \bb{p}_2', \bb{p}_3'} \, \sum_{\bb{p}_1, \bb{p}_2, \bb{p}_3} \, \sum_{\bb{p}_{12}, \bb{p}_{23}, \bb{p}_{31}} \!  \! \delta_{\bb{p}_1',\bb{p}_1 + \bb{p}_{12} - \bb{p}_{31}} \, \delta_{\bb{p}_2',\bb{p}_2 - \bb{p}_{12} + \bb{p}_{23}} \, \delta_{\bb{p}_3',\bb{p}_3 + \bb{p}_{31} - \bb{p}_{23}} V_{\bb{p}_{12}} \, V_{\bb{p}_{23}} \, V_{\bb{p}_{31}} \,  \hat{a}_{\bb{p}_1'}^{\dagger} \hat{a}_{\bb{p}_2'}^{\dagger} \hat{a}_{\bb{p}_3'}^{\dagger} \hat{a}_{\bb{p}_3} \hat{a}_{\bb{p}_2} \hat{a}_{\bb{p}_1 \mbox{,}}
\end{equation}
with
\begin{equation}
\bb{p}_1 + \bb{p}_2 + \bb{p}_3 = \bb{p}_1' + \bb{p}_2' + \bb{p}_3' \label{conservation_law}  \mbox{.}
\end{equation}

Using the relation \eqref{conservation_law} 
and performing the Bogoliubov approximation, it is found that 
\begin{equation}
\begin{aligned}  
\hat{H}_I = & \frac{N_0^3}{6 \, \Omega^3} 
\sum_{\bb{p}_{12}} V_{\bb{p}_{12}}^3  \! + \! \frac{N_0^2}{6 \, \Omega^3}   
\sum_{\bb{p} \neq 0} \sum_{\, \bb{p}_{12}} 
\begin{aligned}[t]
\! \biggl\{ & \Bigl[     3 V_{\bb{p}_{12}}^3 + 
2 V_{\bb{p}_{12}}^2 \, \Bigl( V_{\bb{p}_{12} + \bb{p}}  +  V_{\bb{p}_{12} - \bb{p}}   \Bigr) +  V_{\bb{p}_{12}} \, \Bigl( V_{\bb{p}_{12} + \bb{p}}^2 + V_{\bb{p}_{12} - \bb{p}}^2 \Bigr)  \Bigr]      \hat{a}_{\bb{p}}^{\dagger} \hat{a}_{\bb{p}}  + \\
& + V_{\bb{p}_{12}}^2 \, \Bigl(  V_{\bb{p}_{12} + \bb{p}} + V_{\bb{p}_{12} - \bb{p}} \Bigr) \! \Bigl(   \hat{a}_{\bb{p}}^{\dagger} \hat{a}_{-\bb{p}}^{\dagger} + \hat{a}_{\bb{p}} \hat{a}_{-\bb{p}}   \Bigr) +  \\
& + V_{\bb{p}_{12}} \, \Bigl( V_{\bb{p}_{12} - \bb{p}}^2 \, \hat{a}_{\bb{p}}^{\dagger} \hat{a}_{-\bb{p}}^{\dagger}  + V_{\bb{p}_{12} + \bb{p}}^2  \, \hat{a}_{\bb{p}} \hat{a}_{-\bb{p}}\Bigr)  \biggr\}  \mbox{ .}\label{eqn:Hamiltonian F.T.}
\end{aligned}
\end{aligned}
\end{equation}

To analyse the previous expression for $\hat{H}_I$, 
we denote by $\hat{A}_{\bb{p}}$ 
a general function of the $\hat{a}$'s operators 
entering in \eqref{eqn:Hamiltonian F.T.}. In general 
the terms $V_{\bb{p}_{12}} \, V_{\bb{p}_{12} - \bb{p}}^{2} \hat{A}_{\bb{p}}$ and $V_{\bb{p}_{12}} \, V_{\bb{p}_{12} + \bb{p}}^{2} \hat{A}_{\bb{p}}$ are different. Nonetheless it is possible to show that
\begin{equation}
\sum_{\bb{p} \neq 0} \sum_{\, \, \bb{p}_{12}} V_{\bb{p}_{12}}  V_{\bb{p}_{12} - \bb{p}}^2 \, \hat{A}_{\bb{p}} = \sum_{\bb{p} \neq 0} \sum_{\, \, \bb{p}_{12}} V_{\bb{p}_{12}}  V_{\bb{p}_{12} + \bb{p}}^2 \, \hat{A}_{\bb{p}} \mbox{.} \label{eqn:dummy_rel}
\end{equation}
Indeed, by using the fact that $V_{\bb{p}} = V_{-\bb{p}}$ and by 
doing the change of variables $\tilde{\bb{p}}_{12} = - \bb{p}_{12}$ in 
the left hand side of \eqref{eqn:dummy_rel}, we can rewrite the latter as
\begin{equation}
\sum_{\bb{p} \neq 0} \sum_{\tilde{\bb{p}}_{12} =- \infty}^{+\infty} \! V_{\tilde{\bb{p}}_{12}}  V_{\tilde{\bb{p}}_{12} + \bb{p}}^2 \, \hat{A}_{\bb{p}}  \mbox{,} \label{eqn:change_variables_3}
\end{equation}
so that after relabelling the index 
$\tilde{\bb{p}}_{12} \rightarrow \bb{p}_{12}$, Eq. 
\eqref{eqn:dummy_rel} is proved.

Then the interaction Hamiltonian can be finally written as 
\begin{equation}
\hat{H}_I = \frac{N_T^3}{6 \, \Omega^3} \sum_{\bb{p}_{12}} V_{\bb{p}_{12}}^3  + \frac{N_0^2}{6 \, \Omega^3}   \sum_{\bb{p} \neq 0} \sum_{\, \bb{p}_{12}} \biggl\{  F_{\, \bb{p}_{12},\bb{p}} \, \hat{a}_{\bb{p}}^{\dagger} \hat{a}_{\bb{p}} + \frac{1}{2} F_{\, \bb{p}_{12},\bb{p}}   \Bigl( \hat{a}_{\bb{p}}^{\dagger} \hat{a}_{-\bb{p}}^{\dagger} + \hat{a}_{\bb{p}} \hat{a}_{-\bb{p}}   \Bigr)  \biggr\}  \mbox{,}
\end{equation}
where 
\begin{equation}
F_{\, \bb{p}_{12},\bb{p}} = 2 \Bigl( 2 V_{\bb{p}_{12}}^2  V_{\bb{p}_{12} + \bb{p}} + V_{\bb{p}_{12}}  V_{\bb{p}_{12} + \bb{p}}^2 \Bigr)  \mbox{.}
\end{equation}
The complete Hamiltonian is therefore
\begin{equation}
\hat{H} =  \frac{n^3}{6} \sum_{\bb{p}_{12}} V_{\bb{p}_{12}}^3  +  
\sum_{\substack{ \bb{p} \neq 0  \\ (\bb{p}>0) }} \hat{H}_{\bb{p}} \text{,}
\end{equation}
where
\begin{equation}
\hat{H}_{\bb{p}} = \biggl( \epsilon_p^0 + \frac{n_0^2}{6 \, \Omega} \sum_{\bb{p}_{12}} F_{\, \bb{p}_{12},\bb{p}}  \biggr)   \Bigl( \hat{a}_{\bb{p}}^{\dagger} \hat{a}_{\bb{p}} + \hat{a}_{-\bb{p}}^{\dagger} \hat{a}_{-\bb{p}} \Bigr)  + \frac{n_0^2}{6 \, \Omega} \sum_{\bb{p}_{12}} F_{\, \bb{p}_{12},\bb{p}} \Bigl( \hat{a}_{\bb{p}}^{\dagger} \hat{a}_{-\bb{p}}^{\dagger} + \hat{a}_{\bb{p}} \hat{a}_{-\bb{p}} \Bigr)  \mbox{.}
\end{equation}

The quasi-particle spectrum is therefore
\begin{equation}
\epsilon_{\bb{p}} \! = \! \sqrt{ \! \bigl(\epsilon_p^0\bigr)^{\! 2} \! \! + \! \frac{2}{3} \frac{n_0^2}{\Omega} \biggl( \! 2 \! \sum_{\bb{p}_{12}} \! V_{\bb{p}_{12}}^2  V_{\bb{p}_{12} + \bb{p} } \! + \! \sum_{\bb{p}_{12}} \! V_{\bb{p}_{12}} V_{\bb{p}_{12} + \bb{p}}^2 \! \biggr) \epsilon_p^0 } \mbox{.}
\label{eqn:spectrum_FT}
\end{equation}
We derive Eq. \eqref{eqn:spectrum_FT} in an alternative form in 
Appendix \ref{sec:alternative_derivation_3factors}, where 
we extend to the $3$-body interaction the procedure 
followed in Sec. (\ref{sec:2-body_finite}). 
In Appendix \ref{sec:alternative_derivation_3factors} 
we also show the equivalence of these 
two approaches.

Passing from sums to integrals in Eq. \eqref{eqn:spectrum_FT}, we obtain
\begin{equation}
\epsilon_{\bb{p}} \! = \! \sqrt{ \! \bigl(\epsilon_p^0\bigr)^{\! 2} \! \! + \! \frac{2}{3} \frac{n_0^2}{(2 \pi \hbar)^3} \biggl( \! 2 \! \int \!  \! \mathrm{d}\bb{p}_{12} \, V_{\bb{p}_{12}}^2  V_{\bb{p}_{12} + \bb{p} } \! + \! \int \!  \! \mathrm{d}\bb{p}_{12} \,  V_{\bb{p}_{12}} V_{\bb{p}_{12} + \bb{p}}^2 \! \biggr) \epsilon_p^0 } \mbox{.}
\label{eqn:spectrum_FT_integral}
\end{equation}

\subsection{A specific example of non-local potential} \label{example}

To see how the finite-range in $2$- and $3$-body potentials modifies 
the quasi-particle spectrum, we choose a specific form for it, namely 
a Gaussian form 
\begin{equation}
V(\bb{r}) \propto e^{- \kappa^2 r^2} \mbox{,} \label{eqn:Gaussian_potential}
\end{equation}
applying it to the three cases of $2$-body finite-range 
[Section \ref{sec:2-body_finite}], 
$3$-body finite-range sum of three terms [Section \ref{sub_sub_1}] and 
$3$-body finite-range product of three terms 
[Section \ref{sec:3-body_3-factors}].

For a $2$-body finite-range potential, we put
\begin{equation}
V(\bb{r}) = V \, e^{- \kappa^2 r^2} \mbox{,} \label{eqn:Gaussian_potential_gen}
\end{equation}
with $[V] = [E]$ (and clearly $[\kappa] = [L]^{-1}$). 
The Fourier transform is given by
\begin{equation}
V_{\bb{p}} = \frac{V \pi^{3/2}}{\kappa^3} \, e^{- p^2/4 \hbar^2 \kappa^2} 
\mbox{.} \label{eqn:Gaussian_potential_FT} 
\end{equation}
To make comparison between the $3$-body finite-range potentials we pass to 
dimensionless units, denoted by tildes: we set 
$\tilde{p} =  \frac{p}{2 \, \hbar \, \kappa}$, 
$\tilde{\epsilon}_p =  \frac{\epsilon_p}{\varepsilon}$ and 
$\tilde{V}  = \mathsmaller{\pi^{3/2}} \frac{n_0}{\kappa^3} \frac{V}{\varepsilon}$, 
with $\varepsilon = \frac{2 \, \hbar^2 \, \kappa^2}{m}$. 
In this way the quasi-particle spectrum 
\eqref{eqn:spectrum_2body} can be written as 
\begin{equation}
\tilde{\epsilon}_p = \sqrt{\tilde{p}^4 + 2 \, \tilde{V} e^{-\tilde{p}^2} \tilde{p}^2} \mbox{.} \label{eqn:spectrum1}
\end{equation}

For the $3$-body potential given in Eq. \eqref{eqn:potential_2factors} 
we choose 
\begin{equation}
V(\bb{r}) = {\cal V} 
\, e^{- \kappa^2 r^2} \mbox{,} \label{eqn:Gaussian_potential_gen_3}
\end{equation}
where for the case in consideration $[{\cal V}]=[E]^{1/2}$. One then finds 
\begin{equation}
\tilde{\epsilon}_p = \sqrt{\tilde{p}^4 + 2 \, \tilde{{\cal V}} 
\bigl( 2 \, e^{-\tilde{p}^2} + e^{-2 \tilde{p}^2} \bigr) \tilde{p}^2} \text{,} \label{eqn:spectrum2}
\end{equation}
with $\tilde{{\cal V}} =  \frac{\pi}{3} 
\bigl( \! \frac{n_0}{\kappa^3} \! \bigr)^{\! \mathsmaller{2}} \frac{{\cal V}^2}
{\varepsilon}$ (and $\tilde{p}$ and $\tilde{\epsilon}_p$ defined 
as above).

For the $3$-body potential given in Eq. \eqref{three_fact} 
we choose 
\begin{equation}
V(\bb{r}) = {\cal V}_0 
\, e^{- \kappa^2 r^2} \mbox{,} \label{eqn:Gaussian_potential_gen_3_bis}
\end{equation}
where $[{\cal V}_0]=[E]^{1/3}$. Eq. \eqref{eqn:spectrum_FT_integral} 
assumes then the following simple dimensionless form:
\begin{equation}
\tilde{\epsilon}_p = \sqrt{\tilde{p}^4 + 2 \, \tilde{{\cal V}}_0 
e^{- \frac{2}{3} \tilde{p}^2} \tilde{p}^2} \text{,} \label{eqn:spectrum3}
\end{equation}
with $\tilde{{\cal V}}_0 =  
\frac{\pi}{3 \sqrt{3}} \bigl( \! \frac{n_0}{\kappa^3} 
\! \bigr)^{\! \mathsmaller{2}} \frac{V^3}{\varepsilon}$ and the same definition 
for $\tilde{p}$ and $\tilde{\epsilon}_p$.

\begin{figure}[h]
\begin{center}
\includegraphics[width=.7\columnwidth]{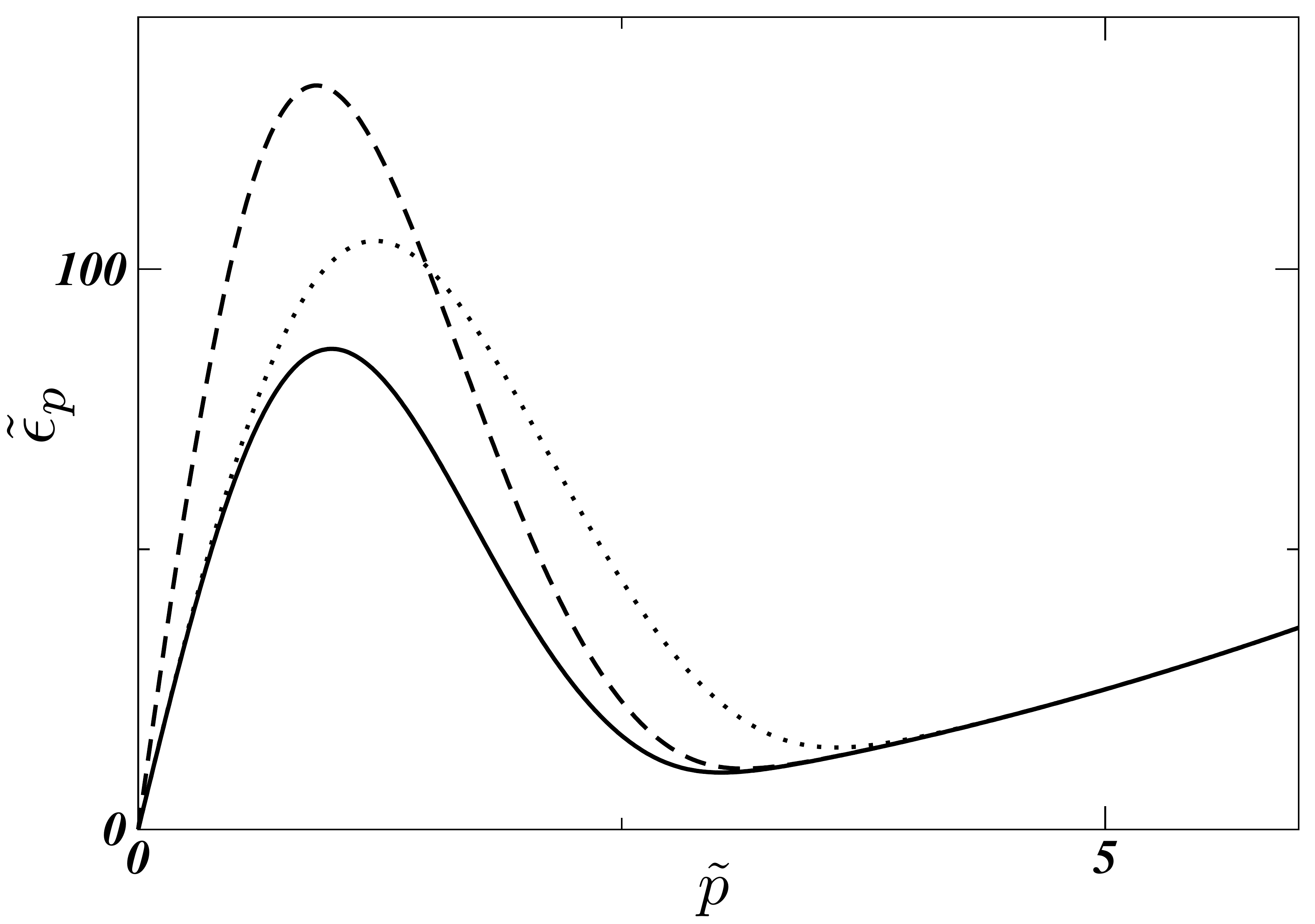}
\end{center}
\caption{
Excitation spectra as given by Eq. \eqref{eqn:spectrum1} 
for the $2$-body finite-range potential \eqref{eqn:Gaussian_potential_gen} 
(solid line), by \eqref{eqn:spectrum2} 
for the $3$-body finite-range potential \eqref{eqn:Gaussian_potential_gen_3} 
(dashed line) 
and by Eq. \eqref{eqn:spectrum3} 
for the $3$-body finite-range potential 
\eqref{eqn:Gaussian_potential_gen_3_bis} (dotted line). Dimensionless units 
are used as explained in Section \ref{example}: in all the three cases 
the dimensionless parameter, respectively $\tilde{V}$, $\tilde{{\cal V}}$ 
and $\tilde{{\cal V}}_0$, is chosen to be $10^4$.}
\label{confronto}
\end{figure}

A comparison between Eqs. \eqref{eqn:spectrum1} and \eqref{eqn:spectrum2} 
shows that the functional form of the quasi-particle spectrum is different 
between $2$- and $3$-body finite-range interactions, and the two 
considered $3$-body finite-range interactions give quite different results. 
To show the differences in the spectra with the same value of 
the dimensionless coupling strength, i.e. setting 
$\tilde{V}=\tilde{{\cal V}}=\tilde{{\cal V}}_0$, 
a plot is presented for the sake of comparison in 
Fig. \ref{confronto}.


%
%

\section{$2$- and $3$-body $\delta$-interactions}
\label{sec:2_3}

As a first application of the results presented in Sections \ref{sec:local} 
and \ref{sec:finite-range} we consider a model with $2$- and $3$-body 
contact interactions and we apply the obtained findings for realistic 
values of the trap parameters.

For a model with $2$- and $3$-body $\delta$-interactions, 
the Hamiltonian reads
\begin{equation}
\hat{H} = \int \!  \mathrm{d}\bb{r} \, \hat{\Psi}^{\dagger}(\bb{r}) \, \biggl( -\frac{\hbar^2 \nabla^2}{2 m}  \biggr) \, \hat{\Psi}(\bb{r})  + 
\frac{U_{2}}{2!}  \int \!  \mathrm{d}\bb{r} \, 
\biggl( \hat{\Psi}^{\dagger}(\bb{r}) \biggr)^{\! 2} 
\cdot \biggl( \hat{\Psi}(\bb{r}) \biggr)^{\! 2} + \frac{U_{3}}{3!}  
\int \!  \mathrm{d}\bb{r} \, 
\biggl( \hat{\Psi}^{\dagger}(\bb{r}) \biggr)^{\! 3} \cdot 
 \biggl( \hat{\Psi}(\bb{r}) \biggr)^{\! 3} \mbox{.} 
\label{eqn:3general_1}
\end{equation}
In the Bogoliubov approximation one finds
\begin{equation}
\hat{H} = \biggl( n \frac{U_2}{2} + n^2 \frac{U_3}{6} \biggr)N_T  + 
\sum_{\substack{ \bb{p} \neq 0  \\ (\bb{p}>0) }}  \biggl\{ \biggl( \epsilon_{p}^0  + X^{(3)} \biggr) \biggl( \hat{a}_{\bb{p}}^{\dagger} \hat{a}_{\bb{p}} + \hat{a}_{-\bb{p}}^{\dagger} \hat{a}_{-\bb{p}} \biggr) + X^{(3)}  \biggl(  \hat{a}_{\bb{p}}^{\dagger} \hat{a}_{-\bb{p}}^{\dagger} + \hat{a}_{\bb{p}} \hat{a}_{-\bb{p}}  \biggr) \biggr\} \mbox{,}
\end{equation}
with
\begin{equation}
X^{(3)} = n_0 U_2 + n_0^2 U_3 \mbox{.}
\end{equation} 
Note that $U_2$ and $U_3$ have dimensions 
$[E] \! \cdot \! [L]^3$ and $[E] \! \cdot \! [L]^6$, respectively. 
The ground-state energy is found to be 
\begin{equation}
\frac{E_0}{\Omega} = \biggl( \frac{n^2 U_2 }{2} + \frac{n^3 U_3}{6} \biggr) + \frac{8}{15 \pi^2} \biggl( \frac{m \, \mathfrak{s}}{\hbar} \biggr)^{\! 3} \mbox{,} \label{eqn:3_ground-state_N3}
\end{equation}
with $\mathfrak{s} = \sqrt{X^{(3)}/m} = 
\sqrt{\bigl( n_0 U_2 + n_0^2 U_3 \bigr)/m}$.

\begin{figure}[t]
\begin{center}
\includegraphics[width=.7\columnwidth]{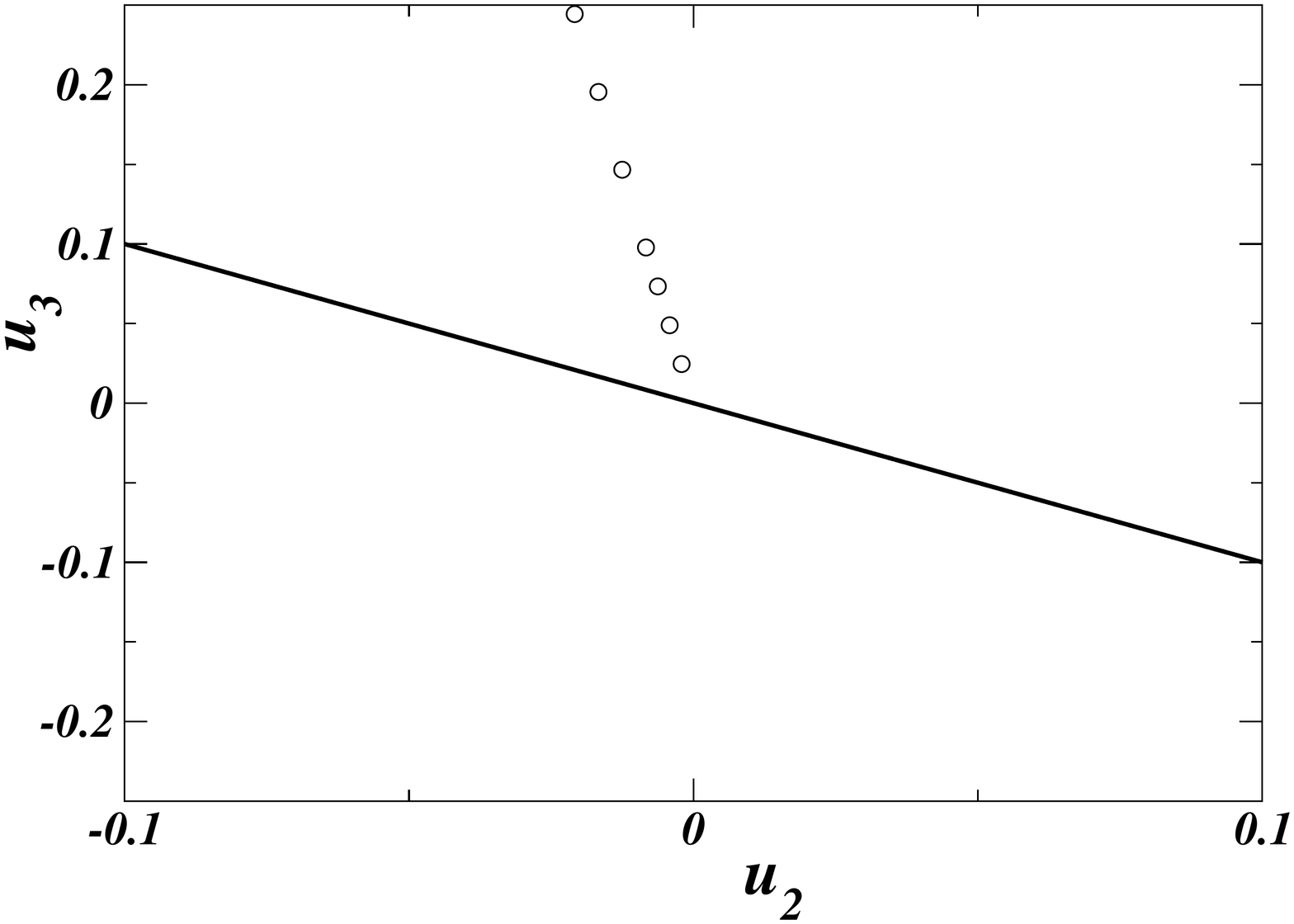}
\end{center}
\caption{
Plot of stability for a Bose gas with $2$- and $3$-body $\delta$-interactions 
according to Eq. \eqref{stab_2_3}.  
Above (below) the separation line the system is stable (unstable). 
We are using the notations $u_2 \equiv U_2/\hbar \bar{\omega} \bar{a}^3$ and 
$u_3 \equiv U_3n_0/\hbar \bar{\omega} \bar{a}^3$. The circles refer 
from the bottom to the values $a_s/a_0=-5,-10,-15,-20,-30,-40,-50 $ with the 
following set of parameters: $N=10000$, $\omega_\perp=2\pi \cdot 100$Hz, 
$\omega_z=2\pi \cdot 10$Hz, $g_3=10^{-27} \hbar \cdot cm \,s^{-1}$. 
The density $n_0$ is estimated as discussed in the text and 
$\bar{\omega}=(\omega_\perp^2 \omega_z)^{1/3}$, 
$\bar{a}=\sqrt{\hbar/m\bar{\omega}}$.}\label{fig:1_stability}
\end{figure}

The stability as a function of $U_1$ and $U_2 / n_0$ is simply given 
by $X^{(3)}>0$, i.e. by
\begin{equation}
U_2 + n_0 U_3 > 0 \mbox{.}
\label{stab_2_3}
\end{equation}  
Using \eqref{stab_2_3} we can evaluate the values 
of $U_3$ for which one has instability, as depicted in 
Fig. \ref{fig:1_stability}. 
To make contact with a notation often used in the literature, 
we set $U_3 \equiv 2g_3$. 
Theoretical estimates for $g_3$ have been given in literature 
\cite{gammal2001,kohler2002,li2010}. 
Regarding one-dimensional trapping potentials, 
we mention that the $3$-body recombination rate in $1D$ Bose gases was  
experimentally studied \cite{laburthe2004,haller2011}, and several 
theoretical studies addressed the problem of determining $g_3$ and 
its effects in the Lieb-Liniger model 
\cite{gangardt2003,kheruntsyan2003,cheianov2006,kormos2009,kormos2011,piroli2015}. 

Finally, the depletion fraction is given by
\begin{equation}
\frac{n_{ex}}{n} = \frac{\sqrt{n}}{3 \pi^2} \biggl( \frac{\sqrt{m (U_2 + n U_3)}}{\hbar} \biggr)^{\! 3} \mbox{,}
\end{equation}
where in the right-hand side it has been used the fact that 
$n_0 \approx n$.

As an application of the previous results we may consider 
the case of $^{85}$Rb atoms, having a negative scattering length $a_s<0$ 
\cite{cornish2000,dorney2001,cornish2006,altin2010,mcdonald2014,everitt2017}. 
In the setup described in \cite{everitt2015} the 
breathing frequency of a $^{85}$Rb gas was studied after a variation of $a_s$.
To refer to a realistic setup and 
reminding that in our study we are not including $3$-body losses 
(so that $g_3=Re[g_3]$), 
we deal with an external potential having the form 
\begin{equation}
V(\vec{r})=\frac{1}{2}m \left[ 
\omega_\perp^2 (x^2+y^2) + \omega_z^2 z^2 \right] \mbox{.}
\label{anis_form}
\end{equation} 
For an anisotropic trap 
of the form \eqref{anis_form} a 
very simple estimate of the density $n_0$ can be obtained 
by setting $n_0 \approx N_T/\Omega$ and choosing as effective volume 
$\Omega$ the product $\ell_x \ell_y \ell_z\equiv \Omega_{TF}$
of the Thomas-Fermi quantities $\ell_\alpha$ (with $\alpha=x,y,z$) defined 
in Appendix \ref{sec:TF_is}, where we study the 
cubic-quintic Gross-Pitaevskii equation 
in an isotropic potential parabolic trap. Similarly to what is done 
for $2$-body interactions, to take into account in a simple way the effect 
of the anisotropy of the potential $V\left( \vec{r} \right)$, 
we use the formula \eqref{omega_TF} of Appendix \ref{sec:TF_is} 
by substituting $\bar{\omega}=(\omega_\perp^2 \omega_z)^{1/3}$ in place 
of $\omega$ and 
$\bar{a}=\sqrt{\hbar/m\bar{\omega}}$ in place of 
$a=\sqrt{\hbar/m\omega}$ [where $\omega$ and $a$ refer to an isotropic 
potential $V\left( \vec{r} \right)=\frac{1}{2} m \omega^2 \left( 
x^2 + y^2 + z^2 \right)$, as studied in Appendix \ref{sec:TF_is}]. 

In Fig.\ref{fig:1_stability}, we report points corresponding to different values 
of $a_s<0$ \cite{cornish2000} for a set of realistic values 
of the parameters, together with result 
\eqref{stab_2_3}. For each value of $a_s$ we determine the 
Thomas-Fermi radius $R$ using Eq. \eqref{TF} and then the quantities 
$\ell_\alpha$ via Eq. \eqref{val_med_r2}. An important 
point to be observed is that when $|a_s|$ increases 
the ratio $\ell_\alpha/R$ increases, but $R$ itself decreases 
(due to the small but appreciable attractive $2$-body interactions), finally 
resulting in a decrease of the effective volume and an increase of 
$U_3n_0$.

To illustrate the results reported in Fig.\ref{fig:1_stability} we introduced 
the dimensionless variable $u_2\equiv
\frac{U_2}{\hbar \bar{\omega} \bar{a}^3}$ and 
$u_3 \equiv \frac{U_3n_0}{\hbar \bar{\omega} \bar{a}^3}$, 
respectively proportional 
to $U_2$ and $U_3$. It emerges from the figure that the points lie 
in the stability region: similar results (even deeper 
in the stability region) would have been obtained if 
we had chosen the peak density $n(0)$ at 
the trap center, which is another reasonable choice. 
We also considered other values of $U_3$ to explore 
the dependence of the stability plot on the values of $U_2$, $U_3$,  and 
the result is that, even with rather larger values of $U_3$,  
the system is stable. We observe that with (the very large value of) 
$a_s=-100a_0$, one gets $|u_3/u_2| \sim 10$. Furthermore, for a 
value $U_3$ which is, e.g., $10$ times larger than the 
one considered in Fig. \ref{fig:1_stability}, 
one would have again $|u_3/u_2| \sim 10$, 
again well inside the stability region.

%

\section{Other applications to ultracold atom systems}
\label{sec:other}

In this Section we discuss some further 
realistic interaction potentials which are of interest for current 
experimental setups with ultracold atoms. 
The discussed applications include dipolar systems, e.g. magnetic
atoms and polar molecules, 
and soft-core potentials that can be simulated with Rydberg dressed atoms.
For each of these cases we analyse the energy of the elementary 
excitations in the homogeneous limit and derive for some interesting
parameter regimes the stability diagram of the Bogoliubov spectra.

Before discussing in detail these specific implementations 
let us consider a general model with a non-local $2$-body 
interaction potential accompanied by $2$- and $3$-body contact interactions. 
Note that it is possible to extend this model straightforwardly 
adding $\delta$-interactions 
up to $N$-body to a $2$-body non-local interaction 
potential, 
but for the sake of simplicity we will not consider this general case.

The Hamiltonian of the system reads
\begin{equation}
\begin{split}
\hat{H} = \int \!  \mathrm{d}\bb{r} \, \hat{\Psi}^{\dagger}(\bb{r}) \, \biggl( \! -\frac{\hbar^2 \nabla^2}{2 m}  \biggr) \, \hat{\Psi}(\bb{r})  + & \frac{1}{2!}  \int \!  \mathrm{d}\bb{r}_1 \mathrm{d}\bb{r}_2 \, \hat{\Psi}^{\dagger}(\bb{r}_1) \hat{\Psi}^{\dagger}(\bb{r}_2) \, V_2(\bb{r}_1-\bb{r}_2) \, \hat{\Psi}(\bb{r}_2) \hat{\Psi}(\bb{r}_1) + \\
+ & \frac{U_{2}}{2!}  \int \!  \mathrm{d}\bb{r} \, 
\biggl( \hat{\Psi}^{\dagger}(\bb{r}) \biggr)^{\! 2} \cdot
\biggl( \hat{\Psi}(\bb{r}) \biggr)^{\! 2}  +  \frac{U_{3}}{3!}  \int \!  \mathrm{d}\bb{r} \, \biggl( \hat{\Psi}^{\dagger}(\bb{r}) \biggr)^{\! 3}  
\cdot \biggl( \hat{\Psi}(\bb{r}) \biggr)^{\! 3} \mbox{.} 
\end{split}
\end{equation}
Proceeding as in Sections \ref{sec:local} and \ref{sec:finite-range} 
and expanding the quantum fields in terms of creation and 
annihilation operators, one arrives in the Bogoliubov approximation 
at the following expression for the Hamiltonian:
\begin{equation}
\hat{H} = \biggl( n \frac{V_0}{2} + n \frac{U_2}{2} + n^2 \frac{U_3}{6} \biggr)N_T  + \sum_{\substack{ \bb{p} \neq 0  \\ (\bb{p}>0) }}  
\biggl\{ \biggl( \epsilon_{p}^0  + X_{\bb{p}} \biggr) \biggl( \hat{a}_{\bb{p}}^{\dagger} \hat{a}_{\bb{p}} + \hat{a}_{-\bb{p}}^{\dagger} \hat{a}_{-\bb{p}} \biggr) + X_{\bb{p}}  \biggl(  \hat{a}_{\bb{p}}^{\dagger} \hat{a}_{-\bb{p}}^{\dagger} + \hat{a}_{\bb{p}} \hat{a}_{-\bb{p}}  \biggr) \biggr\} \mbox{,}
\end{equation}
where $X_{\bb{p}} = n_0 V_{\bb{p}} + n_0 U_2 + n_0^2 U_3$ 
is the sum of the Fourier components of each interaction
potential times the condensate density $n_0$ to the proper power. 
We then arrive at the Bogoliubov excitation spectrum:
\begin{equation}
\epsilon_{\bb{p}} = \sqrt{\bigl(\epsilon_p^0\bigr)^2 + 2 \Bigl( n_0 V_{\bb{p}} + n_0 U_2 + n_0^2 U_3 \Bigr) \, \epsilon_p^0} \mbox{.} \label{eqn:spectrum_isotr}
\end{equation}
The condensate density can be obtained 
by solving the self-consistent equation:
\begin{equation}
n_0 = n - \frac{1}{2(2 \pi \hbar)^3} \int \!  \mathrm{d}\bb{p} \Biggl( \frac{\frac{p^2}{2 m} + n_0 V_{\bb{p}} + n_0 U_2  +  n_0^2 U_3 }{\sqrt{\bigl( \frac{p^2}{2 m} \bigr)^{\! 2} + 2 \bigl( n_0 V_{\bb{p}} + n_0 U_2 + n_0^2 U_3 \bigr) \frac{p^2}{2 m}}} - 1 \Biggr) \mbox{.}
\end{equation}

An example where the stability 
conditions can be carried out explicitly 
is the Gaussian 2-body potential 
Eq. \eqref{eqn:Gaussian_potential_gen} plus a $2$- and $3$-body $\delta$-interaction. 
The Bogoliubov spectrum then reads
\begin{equation} \label{eqn:spectrum_2b_fin_ran+2-3b_delta}
\tilde{\epsilon}_p = \sqrt{ \tilde{p}^4 + 2 (\tilde{V} e^{-\tilde{p}^2} + \tilde{U}_2 + \tilde{U}_3) \tilde{p}^2} \mbox{,}
\end{equation}
with $\tilde{U}_2 = \mathsmaller{n_0} \frac{U_2}{\varepsilon}$, 
$\tilde{U}_3 = \mathsmaller{n_0^2} \frac{U_3}{\varepsilon}$. 
In Eq. \eqref{eqn:spectrum_2b_fin_ran+2-3b_delta} 
dimensionless units are used as in Section \ref{example}, setting  
$\tilde{p} =  \frac{p}{2 \, \hbar \, \kappa}$, 
$\tilde{\epsilon}_p =  \frac{\epsilon_p}{\varepsilon}$ and 
$\tilde{V}  = \mathsmaller{\pi^{3/2}} \frac{n_0}{\kappa^3} \frac{V}{\varepsilon}$, 
with $\varepsilon = \frac{2 \, \hbar^2 \, \kappa^2}{m}$. 

Setting for convenience $\tilde{U} = \tilde{U}_2 + \tilde{U}_3$, 
we easily derive the inequality 
ensuring the stability for each component of the Fourier spectrum:
\begin{equation}
\frac{1}{2 \tilde{V}} \, \tilde{p}^2 + \frac{\tilde U}{\tilde V} > - e^{-p^2} \text{.}
\end{equation}
We then arrive at the following conditions for
the parameters $\tilde{V}$ and $\tilde{U}$:
\begin{equation} \label{eqn:plot_stability_2b_ft+2-3b_d}
\text{Stability:} 
       \begin{cases}
       \tilde{V} > \dfrac{e^{-2 \tilde{U}}}{2 e}  &   \tilde{U} \leq - 1/2 \text{,} \vspace{3mm} \\
       \tilde{V} > -\tilde{U}                            &   \tilde{U} > - 1/2 \mbox{,}
       \end{cases}
\end{equation}
which define the stability regions of the Bogoliubov spectrum.

\begin{figure}[h]
\begin{center}
\includegraphics[width=.7\columnwidth]{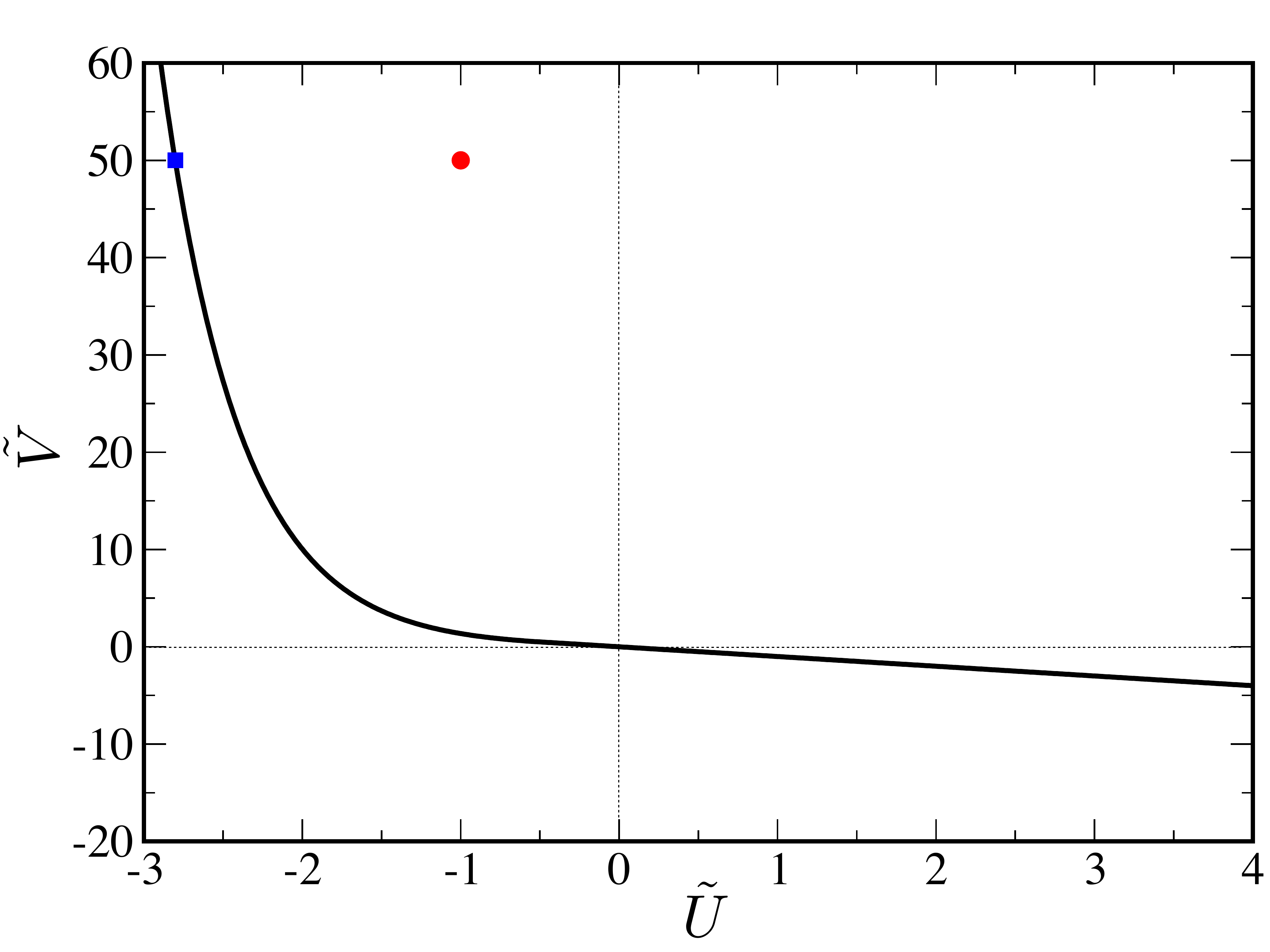}
\end{center}
\caption{
Stability plot for the excitation spectrum 
\eqref{eqn:spectrum_2b_fin_ran+2-3b_delta} as a function of the
dimensionless parameters $\tilde{U}$ and $\tilde{V}$. 
Above (below) the separation line the uniform state is 
stable (unstable) against linear perturbations. 
Notice that competitive interactions or attractive potentials 
are needed to have instability. The red circle and 
the blue square correspond to the values of 
$(\tilde U,\tilde V)$ used in 
Fig. \ref{fig:spectrum_2b_gaussian_and_2-3b_delta}.}
\label{fig:stability_plot_2b_gaussian_and_2-3b_delta}
\end{figure}

\begin{figure}[h]
\begin{center}
\includegraphics[width=.7\columnwidth]{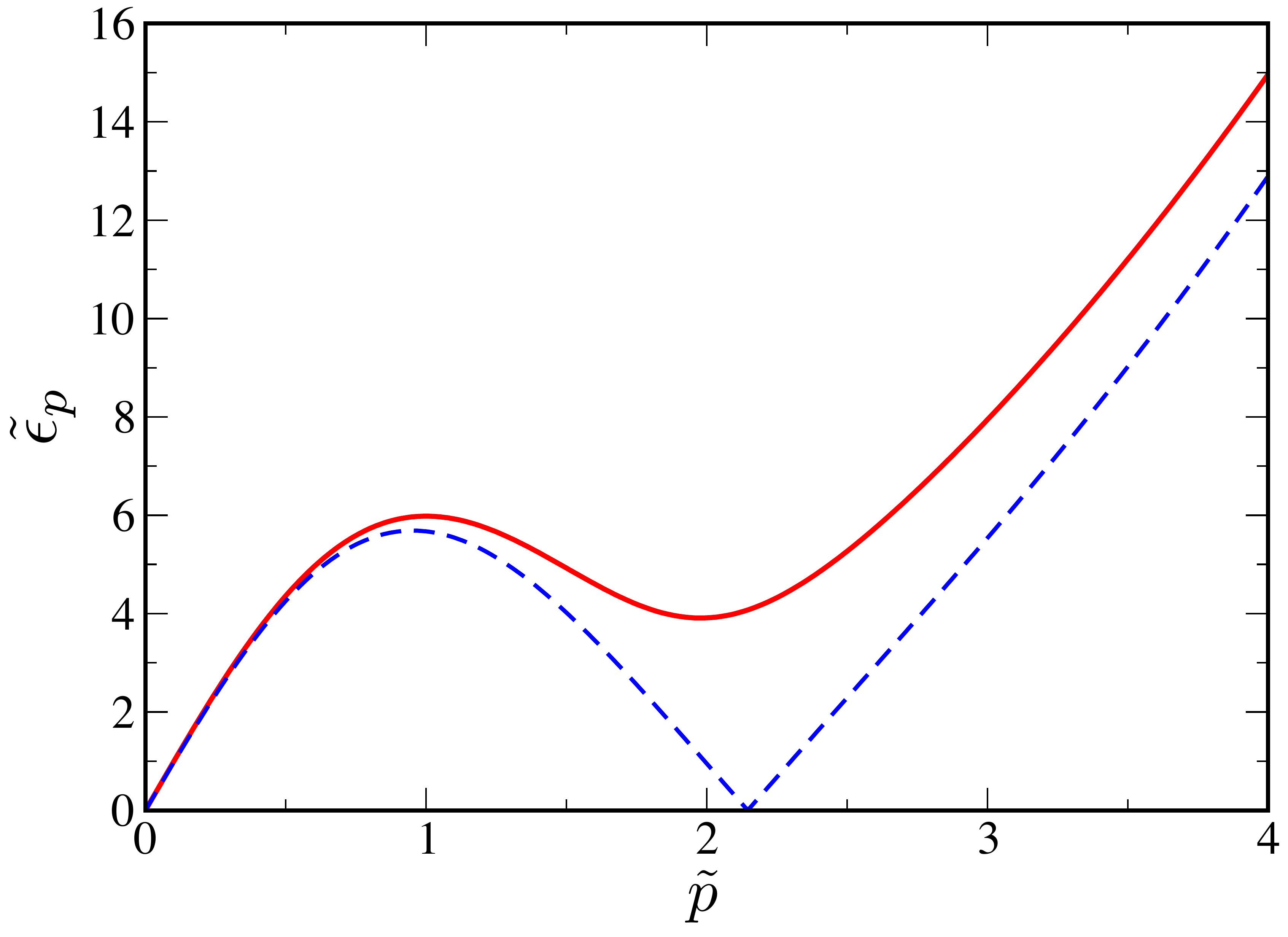}
\end{center}
\caption{
Excitation spectrum \eqref{eqn:spectrum_2b_fin_ran+2-3b_delta} 
with roton-like minima. Red line: 
spectrum for $\tilde V=50$ and $\tilde U=-1$ (corresponding 
to the red circle in 
Fig. \ref{fig:stability_plot_2b_gaussian_and_2-3b_delta}). 
Blue dashed line: spectrum for 
$\tilde V=50$ and $\tilde U=-2.8$, where the 
minimum softens and the uniform solution
becomes unstable signalling the onset of a new modulated ground-state.
This spectrum lies on the separation line
of Fig. \ref{fig:stability_plot_2b_gaussian_and_2-3b_delta} and 
 corresponds to 
the blue square in Fig. \ref{fig:stability_plot_2b_gaussian_and_2-3b_delta}. 
}
\label{fig:spectrum_2b_gaussian_and_2-3b_delta}
\end{figure}

The inequalities \eqref{eqn:plot_stability_2b_ft+2-3b_d} are represented in 
Fig. \ref{fig:stability_plot_2b_gaussian_and_2-3b_delta}. 
Of course, for all repulsive non-local interactions and $2$- and 
$3$-body contact potentials 
the system is stable against linear perturbations around the uniform solution. 
When local and non-local 
potentials are competing (opposite sign) or when both 
$\tilde U$ and $\tilde V$ are negative, it is possible to find instabilities 
which may signal the onset of a structured ground-state configuration.
In Fig. \ref{fig:spectrum_2b_gaussian_and_2-3b_delta} 
we show two cases with negative short-range
interaction $\tilde{U}$ 
where a roton-like minimum occurs. 
The blue dashed line spectrum is unstable to linear perturbations for finite
momentum wave-vector, signalling the onset of a modulated ground-state. 
This spectrum lies on the separation line
of Fig. \ref{fig:stability_plot_2b_gaussian_and_2-3b_delta}.

%

\subsection{$2$-body Rydberg-dressed potentials and $3$-body contact 
interactions}

The results presented above can be applied 
in the case of model potentials like soft-core 
interactions that can be implemented in the laboratory with 
Rydberg-dressed potentials to see the effect of $3$-body terms. 
These potentials were recently created in optical lattices with Rb 
atoms excited to Rydberg states \cite{biedermann16,Zeiher16}. 
The interest for such interactions is general and involves 
the simulation of novel kinds of spin Hamiltonians
\cite{vanBijnen2015,Glaetzle14} for the creation of exotic phases, 
like the supersolid \cite{boninsegni12,cinti14,leonard16,li16},
and for metrological applications \cite{macri16,davis16,gil14,henkel10}. 
Motivated by these experimental results and 
theoretical investigations, we focus on the study of the stability diagram 
of a $2$-body isotropic step-potential:
\begin{equation} \label{soft-pot}
V(r) = 
       \begin{cases}
       C  & r \leq R_0 \text{,} \\
       0    & r > R_0 \mbox{.}
       \end{cases}
\end{equation}
Physically realizable potentials generically display long-range tails, 
decaying generically as a power law ($\sim r^{-3}$ or
$\sim r^{-6}$). However, the model potential of Eq. (\ref{soft-pot}) is 
a good approximation of such more complicated 
real potentials in the sense that the many-body properties found for 
such potential do not differ qualitatively from the 
realistic ones \cite{saccani12,kunimi12,macri13,macri14}. 
It is important to recall that the Gaussian model potential 
described above does not fall in the same 
class of soft-core potentials. The reason is that the Fourier transform of the Gaussian potential never changes sign, 
indicating that such potential alone can never display instabilities as 
the ones that are found for generic soft-core potentials.
In the following we analyse in more detail the $3D$ as well as the 
$2D$ geometries in free space for the
potential \eqref{soft-pot}.

\subsubsection{$3$D Case}

In $3D$, the Fourier transform of Eq. \eqref{soft-pot} is:
\begin{equation}
V_{p} = a \frac{j_1(R_0 p/\hbar)}{R_0 p/\hbar} \mbox{,}
\end{equation}
where $a=4  \pi R_0^3 C$ and $j_1$ 
is the spherical Bessel function of the 1st kind 
$j_1(x) = \frac{\sin{x}}{x^2} \mathsmaller{-} \frac{\cos{x}}{x}$.
From Eq. \eqref{eqn:spectrum_isotr} the excitation spectrum is
\begin{equation}
\epsilon_{\bb{p}} = \sqrt{\bigl(\epsilon_p^0\bigr)^2 + 2 \Bigl(4  \pi \, n_0 R_0^3 \, C  \, \frac{j_1(R_0 p/\hbar)}{R_0 p/\hbar} + n_0 U_2 + n_0^2 U_3 \Bigr) \, \epsilon_p^0} \mbox{.}
\end{equation}
In dimensionless form this excitation spectrum can be written as 
\begin{equation} \label{3dsoft_dimensionless}
\tilde{\epsilon}_p = \sqrt{ \tilde{p}^4 + 2 \Bigl(\tilde{C} \, \frac{j_1(\tilde{p})}{\tilde{p}} + \tilde{U}_2 + \tilde{U}_3\Bigr) \tilde{p}^2} \text{,}
\end{equation}
where we defined 
$\tilde{p} = \frac{R_0 p}{\hbar}$, 
$\varepsilon = \frac{\hbar^4}{2 m R_0^2}$, 
$\tilde{\epsilon}_p = \frac{\epsilon_p}{\varepsilon}$, 
$\tilde{C} = \mathsmaller{4 \pi n_0 R_0^3} \frac{C}{\varepsilon}$, 
$\tilde{U}_2  = \frac{n_0 U_2}{\varepsilon}$ and 
$\tilde{U}_3 = \frac{n_0^2 U_3}{\varepsilon}$. 
Defining also 
$\tilde{U}=\tilde{U}_1 + \tilde{U}_2$, the stability condition is represented by
\begin{equation}
\tilde{p}^2 + 2 \tilde{C} \, \frac{j_1(\tilde{p})}{\tilde{p}} + 2 \, \tilde{U} > 0 \mbox{.}
\end{equation} \label{eqn:inequality_Bessel3D}

\subsubsection{$2D$ Case}

\begin{figure}[h]
\begin{center}
\includegraphics[width=.7\columnwidth]{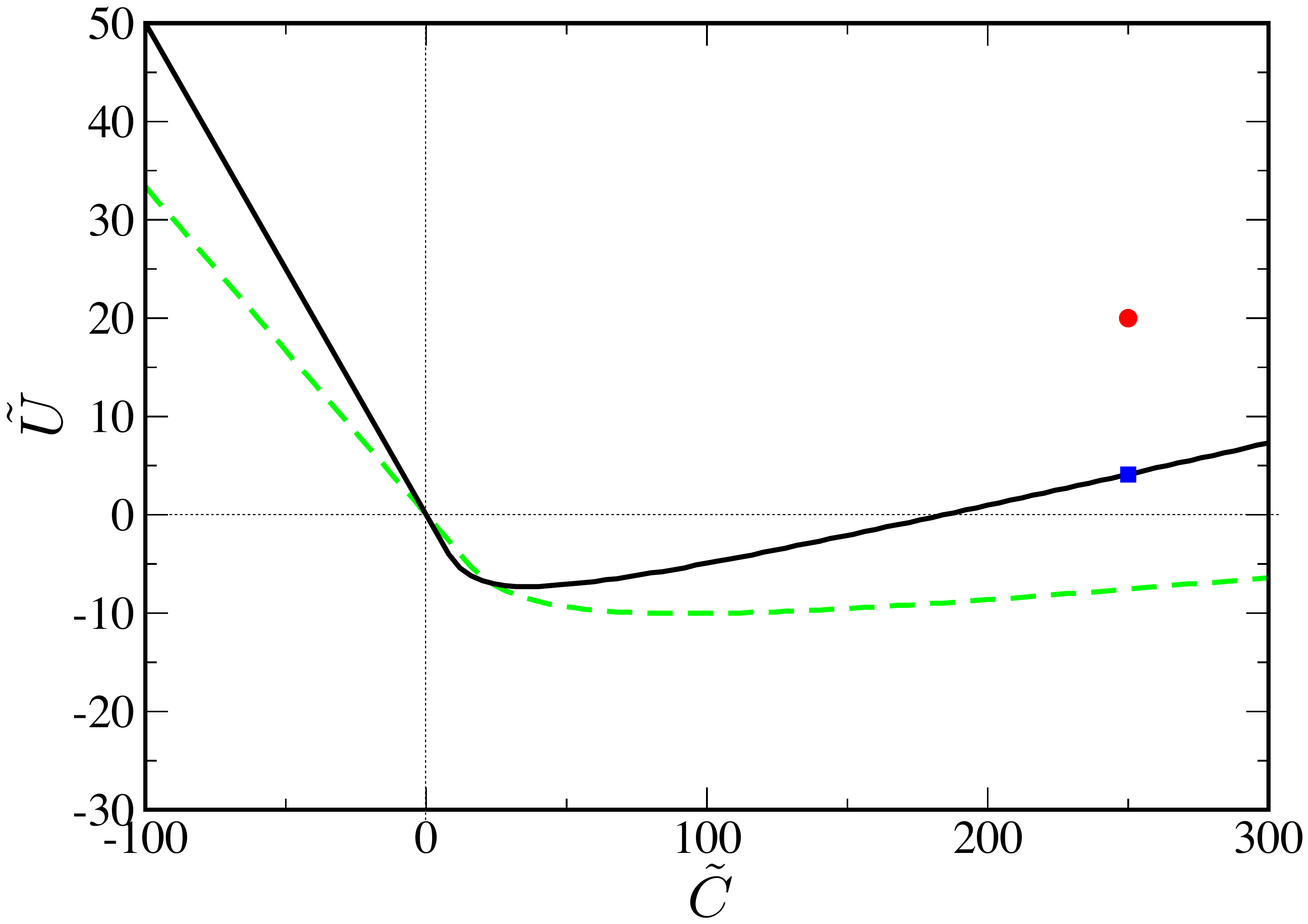}
\end{center}
\caption{
Stability plot for the excitation spectrum 
\eqref{dimless_rydberg} as a function of the
dimensionless parameters $\tilde{U}$ and $\tilde{C}$ for a 
$2D$ soft-core potential. 
Above (below) the separation line the uniform state is 
stable (unstable) against linear perturbations. 
The red circle and the blue square correspond to the 
values $(\tilde{U},\tilde{C})$ of the plots
of Fig. \ref{fig:spectrum_2d_rydberg}.
The green dashed line sets the stability threshold for the $3D$ 
soft-core potential given in Eq. \eqref{3dsoft_dimensionless}. 
Above (below) the separation line the uniform state is 
stable (unstable) against linear perturbations.}
\label{fig:stability_plot_2d_rydberg}
\end{figure}

\begin{figure}[h]
\begin{center}
\includegraphics[width=.7\columnwidth]{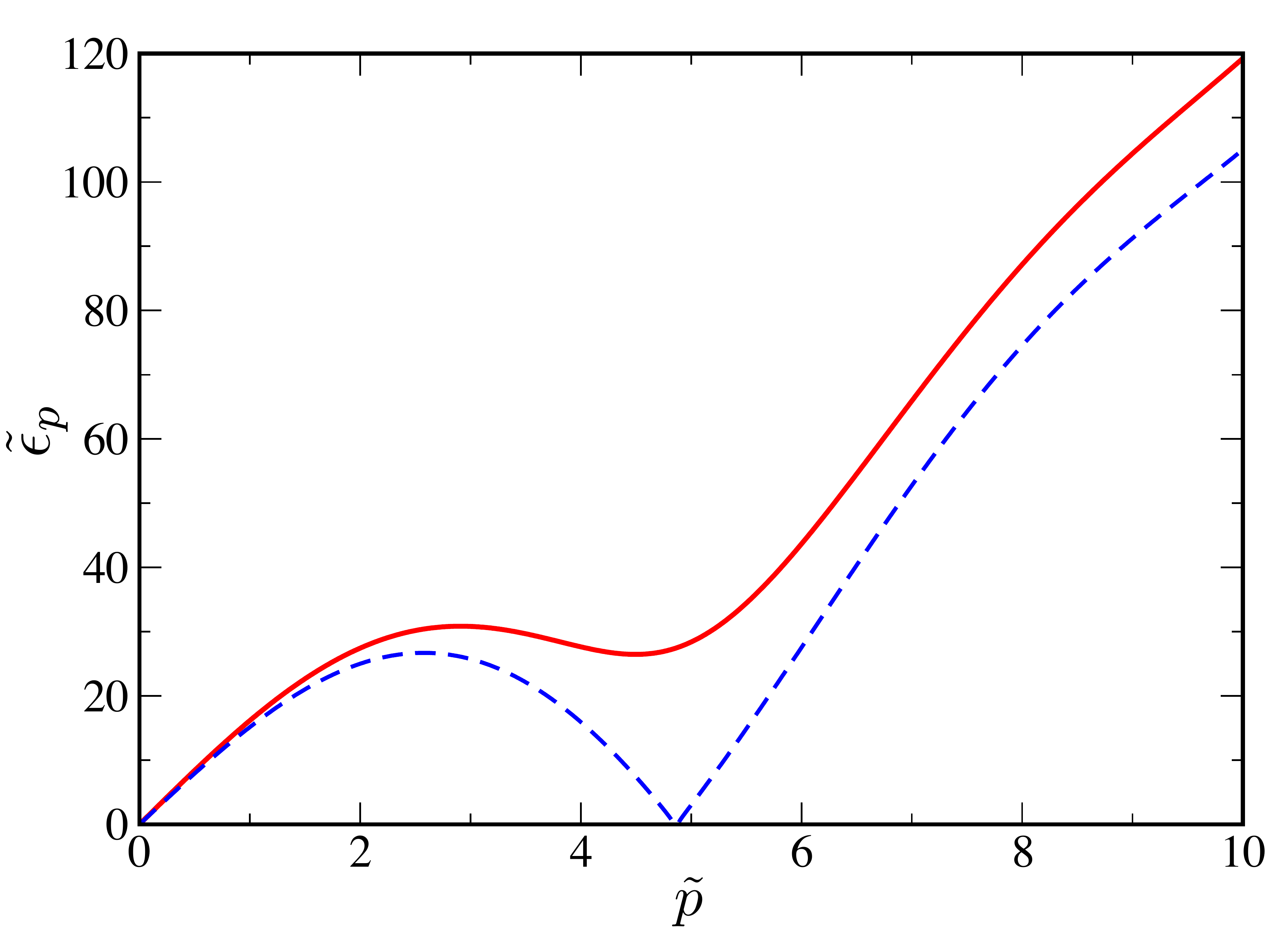}
\end{center}
\caption{
Excitation spectrum \eqref{dimless_rydberg} with rotonic minima. Red line: 
spectrum for $\tilde C=250$ and $\tilde U=20$ (corresponding to the 
red circle in Fig. \ref{fig:stability_plot_2d_rydberg}). 
Blue dashed line corresponding to the blue square in 
Fig. \ref{fig:stability_plot_2d_rydberg}: spectrum for 
$\tilde C=250$ and $\tilde U=4.1$, where the 
minimum softens and the uniform solution
becomes unstable signalling the onset of a new modulated ground-state.
This spectrum lies on the separation line
of Fig. \ref{fig:stability_plot_2d_rydberg}.}
\label{fig:spectrum_2d_rydberg}
\end{figure}

The Fourier transform of Eq. \eqref{soft-pot} in $2D$ is
\begin{equation}
V_{p} = 2  \pi R_0^2 C \frac{J_1(R_0 p/\hbar)}{R_0 p/\hbar} \text{,}
\end{equation}
where $J_1(x)$ is the Bessel function of the 1st kind.
The excitation spectrum is
\begin{equation}
\epsilon_{\bb{p}} = \sqrt{\bigl(\epsilon_p^0\bigr)^2 + 2 \sigma_0 \Bigl(2 \pi R_0^2 C  \frac{J_1(R_0 p/\hbar)}{R_0 p/\hbar} + U_2 + \sigma_0 U_3 \Bigr) \, \epsilon_p^0} \mbox{,}
\end{equation}
with $\sigma_0 \equiv \frac{N_0}{L^2}$. In dimensionless form this expression reads 
\begin{equation} \label{dimless_rydberg}
\tilde{\epsilon}_p = \sqrt{ \tilde{p}^4 + 2 \Bigl(\tilde{C} \frac{J_1(\tilde{p})}{\tilde{p}} + \tilde{U}_2 + \tilde{U}_3\Bigr) \tilde{p}^2} \text{,}
\end{equation}
with 
$\tilde{C} = \mathsmaller{2 \pi \sigma_0 R_0^2} \frac{C}{\varepsilon}$, 
$\tilde{U}_2  = \frac{\sigma_0 U_2}{\varepsilon}$ and  
$\tilde{U}_3 = \frac{\sigma_0^2 U_3}{\varepsilon}$ with 
$\tilde{p}$, $\varepsilon$ and $\tilde{\epsilon}_p$ defined as above. 
Setting $\tilde{U}=\tilde{U}_1 + \tilde{U}_2$, the stability condition is represented by
\begin{equation}
\frac{1}{2 \tilde{C}}\tilde{p}^2 + \frac{\tilde{U}}{\tilde{C}} > - \frac{J_1(\tilde{p})}{\tilde{p}} \mbox{.}
\end{equation}
In Fig. \ref{fig:stability_plot_2d_rydberg} 
we show the stability plots for the $2D$ and $3D$ geometry in terms of the regime parameters $\tilde{C}$ and $\tilde{U}$, picking up two cases for the values $(\tilde{U},\tilde{C})$, represented by a red circle and a blue square; their respective spectra are reported in Fig. \ref{fig:spectrum_2d_rydberg}. 
Notice that, contrarily to the Gaussian potential, in $2D$ 
competitive interactions, or attractive potentials 
are not necessary to have instability. 

%

\subsection{Dipolar interactions in magnetic atoms and $3$-body contact 
interactions}

In this Subsection we analyse the stability diagram of a homogeneous 
bosonic system interacting via a long-range $2$-body dipolar potential.
Such potentials have been investigated in the past years 
for the study of effects induced by non-local interactions in the 
physics of BECs both in free space and in optical lattices 
\cite{lahaye09,baranov12}. 
One of the major problems regarding dipolar interaction 
in free space is their anisotropic character which 
induces instabilities in 
$3D$ homogeneous systems \cite{koch08,lahaye07}. 
On the other hand, the presence of an
asymmetric harmonic trapping in combination with short-range repulsive interactions can eliminate such instabilities,
opening the way to the study of interesting many-body physics with long-range interactions \cite{santos03,fischer06}. 
Recent experiments with dipolar BECs showed that under certain conditions 
where instability is expected
from a standard Bogoliubov approach, dense clusters with many atoms can 
occur \cite{chomaz16,ferrier16,kadau16,schmitt16,ferrier16-1}, 
which are expected to be superfluid \cite{cinti16}.
Two interpretations have been proposed to explain the stabilization of this phase, namely the presence of weak $3$-body interactions \cite{xi16,bisset15} 
and beyond mean-field effects (Lee-Huang-Yang type corrections) 
\cite{baillie16,bisset16,waechtler16}.

Motivated by these recent developments, we analyse in further 
detail the stability of uniform superfluids in the presence
of long-range dipolar interactions and $2$- and $3$-body contact potentials.
The dipolar potential can be written as:
\begin{equation} \label{eqn:anisotropic}
V(\bb{r}) = \frac{C_{dd}}{4 \pi} \frac{1 - 3 \cos^2{\theta}}{r^3}  \mbox{,}
\end{equation}
where $C_{dd}$ is the strength of the dipolar interaction and 
$\theta$ is the angle between the direction of polarization and the 
relative position of the particles. 
Tuning the relative angle among the particles and the 
quantization axis the potential can be either attractive and repulsive. 
The Fourier transform of Eq. \eqref{eqn:anisotropic} is 
\begin{equation}
V_{\bb{p}} = C_{dd} \Bigl( \cos^2{\alpha} - \frac{1}{3} \Bigr) \mbox{,}
\end{equation}
where $\alpha$ is the angle between $\bb{p}$ 
and the polarization direction (of the dipole-dipole interaction).
 
The excitation spectrum is then found to be:
\begin{equation}
\epsilon_{\bb{p}} = \sqrt{\bigl(\epsilon_p^0\bigr)^2 \! + 2 \Bigl[ n_0 \, C_{dd} \Bigl( \cos^2{\alpha} - \frac{1}{3} \Bigr) \! + n_0 U_2 + n_0^2 \, U_3 \Bigr] \, \epsilon_p^0} \mbox{,} \label{eqn:anisotropic_spectrum1}
\end{equation}
with condensate density given by the equation
\begin{equation}
n_0 = n - \frac{1}{2(2 \pi \hbar)^3} \int \!  \mathrm{d}\bb{p} \Biggl( \frac{\frac{p^2}{2 m} + n_0 \, C_{dd} ( \cos^2{\alpha} - \frac{1}{3} ) \! + n_0 U_2 + n_0^2 \, U_3 }{\sqrt{\bigl( \frac{p^2}{2 m} \bigr)^{\! 2} + 2 \bigl[ n_0 \, C_{dd} ( \cos^2{\alpha} - \frac{1}{3} ) \! + n_0 U_2 + n_0^2 \, U_3 \bigr] \frac{p^2}{2 m}}} - 1 \Biggr) \mbox{.}
\end{equation}
The spectrum \eqref{eqn:anisotropic_spectrum1} 
can be expressed in terms of the ratio
\begin{equation}
\varepsilon_{dd} =\frac{a_{dd}}{a_s} = \frac{C_{dd}}{3 U_2}
\end{equation}
of the dipolar length $a_{dd} = \frac{C_{dd} \, m}{12 \pi \hbar^2}$ 
to the $s$-wave scattering length, which compares the relative strength 
of the dipolar and contact interactions:
\begin{equation}
\epsilon_{\bb{p}} = \sqrt{\bigl(\epsilon_p^0\bigr)^2 \! + 2 \, n_0 \Bigl[ 3 \, \varepsilon_{dd} \, U_2 \Bigl( \cos^2{\alpha} - \frac{1}{3} \Bigr) \! + U_2 + n_0 \, U_3 \Bigr] \, \epsilon_p^0} \mbox{.} \label{eqn:anisotropic_spectrum}
\end{equation}
In dimensionless units one gets
\begin{equation} \label{dip_dimensionless}
\tilde{\epsilon}_p = \sqrt{ \tilde{p}^4 + 2 \Bigl[  \Bigl(3 \, \varepsilon_{dd} \cos^2{\alpha} - \varepsilon_{dd} + 1 \Bigr) \tilde{U}_2 + \tilde{U}_3 \Bigr] \tilde{p}^2} \text{,}
\end{equation}
where $\tilde{p} = \frac{p \, a_{dd}}{\hbar}$, 
$\tilde{\epsilon}_p = \frac{\epsilon_p}{\varepsilon}$, 
$\tilde{U}_2 = \frac{n_0 U_2}{\varepsilon}$ and 
$\tilde{U}_3 = \frac{n_0^2 U_3}{\varepsilon}$, with 
$\varepsilon=\frac{\hbar^2}{2 m a_{dd}^2}$.

\begin{figure}[h]
\begin{center}
\includegraphics[width=.7\columnwidth]{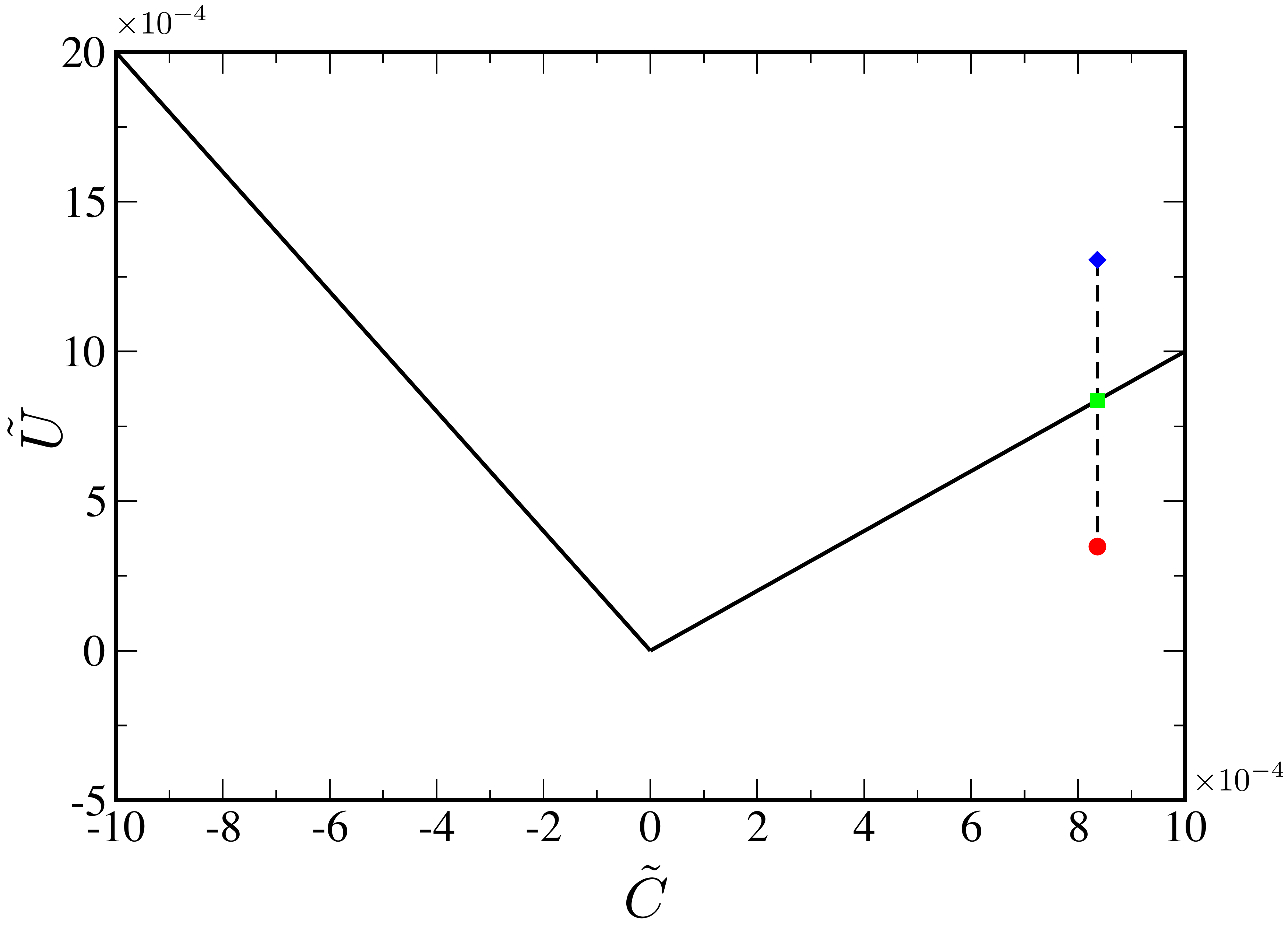}
\end{center}
\caption{
{Stability plot for the excitation spectrum \eqref{dip_dimensionless} 
as a function of the dimensionless parameters 
$\tilde U= \tilde U_2+ \tilde U_3$ and 
$\tilde C=\varepsilon_{dd} \tilde U_2$ for $\alpha=\pi/2$ 
for a $3D$ dipolar potential and contact potentials. 
Above (below) the separation line the uniform state is 
stable (unstable) against linear perturbations.
Data points correspond to three different values of the scattering length for $^{164}$Dy atoms:
$a_s=50\,a_0$ (red circle), $a_s=126\,a_0$ (green square), 
and $a_s=200\,a_0$ (blue diamond), where $a_0$
is the Bohr radius.
In the calculations we included a $3$-body contact interaction potential using 
the value $U_3=5.87\cdot 10^{-27}\, \hbar\, cm^6\,s^{-1}$ considered 
in \cite{bisset15}, 
which slightly enhances the stability region of 
the uniform phase to $\varepsilon_{dd}<1.05$ (green square). 
More quantitatively, putting $U_3=0$ one would have the green square on the 
stability line for $a_s=a_{dd}=131\,a_0$. 
Here we set the average density $n_0=2\cdot 10^{20}\, m^{-3}$ 
and the dipolar length $a_{dd}=131\,a_0$.
}}
\label{fig:stability_plot_3d_magnetic}
\end{figure}

In Fig. \ref{fig:stability_plot_3d_magnetic} 
we plot the stability diagram of Eq. \eqref{dip_dimensionless} 
as a function of the 
dimensionless parameters $\tilde U = \tilde U_2+ \tilde U_3$ and 
$\tilde C = \varepsilon_{dd} \tilde U_2$ for $\alpha=\pi/2$.
We compute the specific values of these parameters for a 
condensate of $^{164}$Dy
also considering as the $3$-body contact interaction the value given in 
\cite{bisset15} (notice however that by using other 
values of $U_3$, as the ones given in \cite{waechtler16}, the effect 
of such terms is anyway rather small). 
From Fig. \ref{fig:stability_plot_3d_magnetic} one sees 
that upon varying the scattering length 
the uniform phase goes from a stable (blue 
diamond) to an unstable (red circle) configuration.
Note that the $3$-body interaction enhances the stability 
region to values of $\varepsilon_{dd}$ larger than $1$.

%

\section{Conclusions}\label{sec:concl}

In this work we have presented a systematic study of 
weakly interacting bosonic gases with local and 
non-local multi-body interactions in the Bogoliubov approximation. 
We considered conservative multi-body interactions for which the number 
of particles is conserved. In fact multi-body interactions 
are associated with the presence of 
particle losses \cite{kagan1985}, that we did not study, rather focusing 
on the determination of the stability conditions due to the 
competition between $2$- and higher-body interactions.
 
A variety of interparticle potentials have been considered. 
We first considered contact interactions, studying the case 
in which the interparticle potential can be written 
as a general sum of $N$-body $\delta$-interactions, providing 
the quasi-particle spectrum, the ground-state energy 
and the stability conditions. Results for general effective 
contact potentials are also presented. 
Our findings show that 
the well-known results for the $2$-body $\delta$-interactions 
in the homogeneous case are generalized in the Bogoliubov 
approximation by the substitution $U_2 n_0 \to X$, where 
$X$ is a function of the condensate fraction $n_0$ 
given by Eq. \eqref{eqn:definition_X_N} for potentials which are sums 
of $N$-body $\delta$-potentials and by Eq. \eqref{eqn:definition_X_N_general} 
in the general case. 
Since the Bogoliubov approximation 
works well when $n_0 \approx n$, 
then one can make the substitution $n_0 \to n$ in $X$, resulting 
for a sum of $N$-body contact interactions 
in the substitution 
$U_2 n \to \sum_{\ell} U_{\ell} \ell (\ell-1) n^{\ell-1}/ \ell!$. 
The case of  higher-body non-local interactions is instead different 
from this respect and the final 
results depend on the specific form of the interactions. 
We explicitly considered two different cases of $3$-body 
non-local interactions. 

In the last part we discussed a few interaction 
potentials which are of interest for current 
experimental setups with ultracold atoms. Implementations include 
systems with $2$- and $3$-body $\delta$-interactions, where we applied 
in the homogeneous limit for realistic values of the trap parameters. 
We also considered the effect of (conservative) $3$-body 
terms in dipolar systems, e.g. magnetic
atoms and polar molecules, and soft-core potentials 
that can be simulated with Rydberg dressed atoms.
For each of these cases we analysed the energy of the elementary 
excitations and derived the stability diagram of the Bogoliubov spectra
for some interesting parameter regimes.

In the present paper we focused on 
higher-body interactions in the homogeneous limit, having in mind 
both $3$-body terms and general effective multi-body interactions. 
Of course ultracold experiments are done in confined traps, 
and we think that a systematic study of the Bogoliubov equations 
in inhomogeneous potentials with general multi-body 
local and non-local interactions is an interesting direction of future research.

\section*{Acknowledgements}

Discussions with L. Barbiero, G. Gori and L. Salasnich are gratefully 
acknowledged. Useful correspondence with N. Robins is as well acknowledged. 
T.M. acknowledges CNPq for support through Bolsa 
de produtividade em Pesquisa n. 311079/2015-6 and  the hospitality of the Physics Department of the University of Padova. 
Support form the European STREP 
MatterWave is acknowledged.

\appendix

\section{Bogoliubov approximation for a general contact interaction} 
\label{sec:spectrum_general_F}

In this Appendix we consider the case of a general contact interaction described by a Hamiltonian of the form  
\begin{equation}
\hat{H} = \hat{H}_0 + \hat{H}_I = \int \!  \mathrm{d}\bb{r} \, \hat{\Psi}^{\dagger}(\bb{r}) \, \biggl( -\frac{\hbar^2 \nabla^2}{2 m}  \biggr) \, \hat{\Psi}(\bb{r})   + 
\int \!  \mathrm{d}\bb{r} \, \colon  \! {\cal F} \! \left( \hat{\rho} \right) \! \colon \mbox{.} 
\end{equation}
One has
\begin{equation}
\begin{aligned}
\hat{H}_I = \int \!  \mathrm{d}\bb{r} \, \colon \!
{\cal F} \! \left( \hat{\rho} \right) \! \colon 
&= \int \!  \mathrm{d}\bb{r} \, \colon \! {\cal F}  \Biggl(  \frac{1}{\Omega} \sum_{\bb{p},\bb{p'}}  e^{-\frac{i}{\hbar} (\bb{p}' \! -\bb{p}) \cdot \bb{r}} \,  \hat{a}_{\bb{p'}}^{\dagger} \hat{a}_{\bb{p}} \Biggr) \! \colon  \\ 
&= \int \!  \mathrm{d}\bb{r} \, \colon \! {\cal F} \Biggl(  \frac{N_0}{\Omega} + \frac{\sqrt{N_0}}{\Omega} \sum_{\bb{p} \neq 0} \biggl( e^{-\frac{i}{\hbar} \bb{p} \cdot \bb{r}} \,  \hat{a}_{\bb{p}}^{\dagger} + e^{\frac{i}{\hbar} \bb{p} \cdot \bb{r}} \, \hat{a}_{\bb{p}} \biggr) + \frac{1}{\Omega} \sum_{\bb{p},\bb{p'} \neq 0}  e^{-\frac{i}{\hbar} (\bb{p}'\!-\bb{p}) \cdot \bb{r}} \,  \hat{a}_{\bb{p'}}^{\dagger} \hat{a}_{\bb{p}} \Biggr) \! \colon \text{,}
\end{aligned}
\end{equation}
where we wrote explicitly the operator $\hat{\rho}$ in terms of the operators $\{ \hat{a}_{\bb{p}}$, $ \hat{a}_{\bb{p}}^{\dagger} \}$ and we used the Bogoliubov approximation. We assume that the function  ${\cal F}={\cal F}(x)$ can be expanded in series up to the second order.

Since $N_0 \approx N_T$ and $N_T \gg 1$, it follows that ${N_0} \gg \sqrt{N_0} \gg 1$, we can write:
\begin{equation}
{\cal F} \! \left( \hat{\rho} \right) \simeq {\cal F} ( n_0 ) + \frac{\partial {\cal F} \! \left( x \right)}{\partial x}\bigg\vert_{x=n_0} \! \! \! \! \! \! \! \! \cdot \, \,\Gamma + \frac{1}{2!} \frac{\partial^2 {\cal F} \! \left( x \right)}{\partial x^2}\bigg\vert_{x=n_0}  \! \! \! \! \! \! \! \! \cdot \, \, \Gamma^{2} \, + \cdots \mbox{,}
\end{equation}
where
\begin{equation}
\Gamma = \sqrt{\frac{n_0}{\Omega}} \sum_{\bb{p} \neq 0} \biggl( e^{-\frac{i}{\hbar} \bb{p} \cdot \bb{r}} \,  \hat{a}_{\bb{p}}^{\dagger} + e^{\frac{i}{\hbar} \bb{p} \cdot \bb{r}} \, \hat{a}_{\bb{p}} \biggr) + \frac{1}{\Omega} \sum_{\bb{p},\bb{p'} \neq 0}  e^{-\frac{i}{\hbar}(\bb{p}'\!-\bb{p}) \cdot \bb{r}} \,  \hat{a}_{\bb{p'}}^{\dagger} \hat{a}_{\bb{p}} \mbox{.}
\end{equation}

Neglecting products of 3 or more operators, the interaction part reads
\begin{equation}
\begin{aligned}
\hat{H}_I 
&= \int \!  \mathrm{d}\bb{r} \, \colon \! \! 
\begin{aligned}[t] 
\biggl\{ &
{\cal F} ( n_0)   
+ \frac{\partial {\cal F} \! \left( x \right)}{\partial x}\bigg\vert_{x=n_0}  \!  
\biggl( \! \sqrt{\frac{n_0}{\Omega}} \sum_{\bb{p} \neq 0} \Bigl( e^{-\frac{i}{\hbar} \bb{p} \cdot \bb{r}} \,  \hat{a}_{\bb{p}}^{\dagger} + e^{\frac{i}{\hbar} \bb{p} \cdot \bb{r}} \, \hat{a}_{\bb{p}} \Bigr) + \frac{1}{\Omega} \sum_{\bb{p},\bb{p'} \neq 0}  e^{-\frac{i}{\hbar} (\bb{p}'\!-\bb{p}) \cdot \bb{r}} \,  \hat{a}_{\bb{p'}}^{\dagger} \hat{a}_{\bb{p}} \biggr) + \\
&+ \frac{1}{2!} \frac{\partial^2 {\cal F} \! \left( x \right)}{\partial x^2}\bigg\vert_{x=n_0}  \!   
\biggl( \frac{n_0}{\Omega} \! \sum_{\bb{p},\bb{p'} \neq 0} \Bigl( e^{-\frac{i}{\hbar} \bb{p} \cdot \bb{r}} \,  \hat{a}_{\bb{p}}^{\dagger} + e^{\frac{i}{\hbar} \bb{p} \cdot \bb{r}} \, \hat{a}_{\bb{p}} \Bigr) \Bigl( e^{-\frac{i}{\hbar} \bb{p}' \! \cdot \bb{r}} \,  \hat{a}_{\bb{p}'}^{\dagger} + e^{\frac{i}{\hbar} \bb{p}' \! \cdot \bb{r}} \, \hat{a}_{\bb{p}'} \Bigr) \! \biggr) 
\! \biggr\} \colon  
\end{aligned}\\ 
&= 
{\cal F} ( n_0 ) \, \Omega  
\begin{aligned}[t] 
&
+ \frac{\partial {\cal F} \! \left( x \right)}{\partial x}\bigg\vert_{x=n_0}  \sum_{\bb{p} \neq 0}  \hat{a}_{\bb{p}}^{\dagger} \hat{a}_{\bb{p}} + \\ 
&+ \frac{1}{2!} n_0  \frac{\partial^2 {\cal F} \! \left( x \right)}{\partial x^2}\bigg\vert_{x=n_0}     
   \sum_{\bb{p} \neq 0} \Bigl(  2  \hat{a}_{\bb{p}}^{\dagger}   \hat{a}_{\bb{p}} +  \hat{a}_{\bb{p}}^{\dagger}\hat{a}_{-\bb{p}}^{\dagger} +  \hat{a}_{\bb{p}} \hat{a}_{-\bb{p}}  \Bigr)  \text{,}
\end{aligned} 
\end{aligned}
\end{equation}
where we integrated out the space variable and used the conservation of momentum, having taken explicitly the normal ordering of the operators.
Using the conservation of the total number of particles, we can write
\begin{equation}
{\cal F} ( n_0 ) \simeq {\cal F} ( n ) - \frac{\partial {\cal F} \! \left( x \right)}{\partial x}\bigg\vert_{x=n} \frac{1}{\Omega} \sum_{\bb{p} \neq 0}\hat{a}_{\bb{p}}^{\dagger}   \hat{a}_{\bb{p}}  \text{,}
\end{equation}
so that 
\begin{equation}
\begin{aligned}
\hat{H}_I  
&=
{\cal F} ( n ) \, \Omega  
\begin{aligned}[t] 
&
+ \biggl( \frac{\partial {\cal F} \! \left( x \right)}{\partial x}\bigg\vert_{x=n_0} \! \! \! \! \! \! \! - \, \, \, \frac{\partial {\cal F} \! \left( x \right)}{\partial x}\bigg\vert_{x=n} \biggr) \! \sum_{\bb{p} \neq 0}  \hat{a}_{\bb{p}}^{\dagger} \hat{a}_{\bb{p}} + \\ 
&+ \frac{1}{2!} n_0  \frac{\partial^2 {\cal F} \! \left( x \right)}{\partial x^2}\bigg\vert_{x=n_0}     
   \sum_{\bb{p} \neq 0} \Bigl(  2  \hat{a}_{\bb{p}}^{\dagger}   \hat{a}_{\bb{p}} +  \hat{a}_{\bb{p}}^{\dagger}\hat{a}_{-\bb{p}}^{\dagger} +  \hat{a}_{\bb{p}} \hat{a}_{-\bb{p}}  \Bigr)  \text{.}
\end{aligned}
\end{aligned}
\end{equation}

Denoting ${\cal G}(y) \equiv \frac{\partial{\cal F} \left( x \right)}{\partial x}\Big\vert_{x=y}$, we see that 
\begin{equation}
{\cal G} (n_0) - {\cal G} (n) \simeq - \frac{\partial {\cal G}(x)}{\partial x}\bigg\vert_{x=n} \frac{1}{\Omega} \sum_{\bb{p} \neq 0}  \hat{a}_{\bb{p}}^{\dagger} \hat{a}_{\bb{p}} \text{,}
\end{equation} 
so that the difference in the second term of the previous equation is of higher order and it can be safely neglected.
Thus the complete Hamiltonian can be written as
\begin{equation}
\hat{H} =  {\cal F} ( n ) \, \Omega + \! \sum_{\substack{ \bb{p} \neq 0  \\ (\bb{p}>0) }} \!  \biggl\{ \! \biggl( \! \epsilon_{p}^0 \! + \! n_0  \frac{\partial^2 {\cal F} \! \left( x \right)}{\partial x^2}\bigg\vert_{x=n_0} \biggr) \! \Bigl( \hat{a}_{\bb{p}}^{\dagger} \hat{a}_{\bb{p}} + \hat{a}_{-\bb{p}}^{\dagger} \hat{a}_{-\bb{p}}  \Bigr) + n_0  \frac{\partial^2 {\cal F} \! \left( x \right)}{\partial x^2}\bigg\vert_{x=n_0} \! \! \Bigl( \hat{a}_{\bb{p}}^{\dagger} \hat{a}_{-\bb{p}}^{\dagger} + \hat{a}_{\bb{p}} \hat{a}_{-\bb{p}}  \Bigr) \! \biggr\} \mbox{,} 
\end{equation}
from which Eq. \eqref{eqn:H_general_X} follows.

\section{Excitation spectrum for a $3$-body non-local factorizable 
potential} 
\label{sec:alternative_derivation_3factors}

In this Appendix we consider the $3$-body finite-range potential 
of Eq. \eqref{three_fact}:
\begin{equation}
U(\bb{r}_1,\bb{r}_2,\bb{r}_3) = V(\bb{r}_1 -\bb{r}_2)V(\bb{r}_2 - \bb{r}_3)V(\bb{r}_3 - \bb{r}_1) \mbox{,}
\end{equation}
where $V(\bb{r}) = V(r)$. In momentum space, the Hamiltonian 
can be written as $\hat{H} = \hat{H_0} + \hat{H}_I$ 
with the kinetic term $\hat{H_0}$ given by Eq. \eqref{kin_Ham}. 
To write the interaction term $\hat{H}_I$, we consider here 
the change of variables 
\{$\bb{r}_1, \bb{r}_2, \bb{r}_3$\} $\rightarrow$ \{$\bb{r}_{12}, \bb{r}_{31}, 
\bb{R}$\}, where
\begin{equation}
\begin{split}
&\bb{r}_{12} = \bb{r}_1 - \bb{r}_2 \mbox{,} \\
&\bb{r}_{31} = \bb{r}_3 - \bb{r}_1 \mbox{,} \\
&\bb{R} = \frac{\bb{r}_1 + \bb{r}_2 +\bb{r}_3}{3} \mbox{.}
\end{split}
\end{equation} 

In the Bogoliubov approximation, after some algebra we obtain
\begin{equation}
\begin{aligned}
\hat{H}_I =& \frac{N_0^3}{6 \Omega^2} \! \int \!  \mathrm{d}\bb{r}_{12} \, \mathrm{d}\bb{r}_{31}  \, V(r_{12}) V(r_{31}) V(r_{12} \! + \! r_{31}) + \\
&+ \frac{N_0^2}{6 \Omega^2} \sum_{\bb{p} \neq 0} 
\begin{aligned}[t]  
\biggl\{ & \int \!  \mathrm{d}\bb{r}_{12} \, \mathrm{d}\bb{r}_{31} \,  \mathfrak{F}_{\bb{p}}(\bb{r}_{12},\bb{r}_{31}) V(r_{12}) V(r_{31}) V(r_{12} \! + \! r_{31}) \Bigl( \hat{a}_{\bb{p}}^{\dagger} \hat{a}_{-\bb{p}}^{\dagger} \! + \! \hat{a}_{\bb{p}} \hat{a}_{-\bb{p}} \Bigr) \! + \\
&\! + \! \! \int \!  \mathrm{d}\bb{r}_{12} \, \mathrm{d}\bb{r}_{31} \Bigl[ 3 + 2 \, \mathfrak{F}_{\bb{p}}(\bb{r}_{12},\bb{r}_{31}) \Bigr]  V(r_{12}) V(r_{31}) V(r_{12} \! + \! r_{31}) \hat{a}_{\bb{p}}^{\dagger} \hat{a}_{\bb{p}}  \biggr\} \mbox{,}
\end{aligned}
\end{aligned}
\end{equation}
where we defined
\begin{equation}
\mathfrak{F}_{\bb{p}}(\bb{r}_{12},\bb{r}_{31}) = \Bigl(  e^{-\frac{i}{\hbar}\bb{p} \cdot \bb{r}_{12}} + e^{-\frac{i}{\hbar}\bb{p} \cdot \bb{r}_{31}} + e^{-\frac{i}{\hbar}\bb{p} \cdot (\bb{r}_{12} + \bb{r}_{31})} \Bigr) \mbox{.}
\end{equation}

Setting
\begin{equation}
C = \int \!  \mathrm{d}\bb{r}_{12} \, \mathrm{d}\bb{r}_{31}  \, V(r_{12}) V(r_{13}) V(r_{12} + r_{13}) \mbox{,}
\end{equation}
proceeding as in Section \ref{sec:2-body_finite}, one can write 
the Hamiltonian in the form
\begin{equation}
\begin{aligned}
\hat{H}   = \frac{N_T^3}{6 \Omega^2}C \! + \! \! \sum_{\bb{p} > 0} 
\begin{aligned}[t]
\biggl\{ & \biggl[ \epsilon_p^0 \! + \! \frac{n_0^2}{3} \! \! \int \! \! \mathrm{d}\bb{r}_{12} \, \mathrm{d}\bb{r}_{31} \,  \mathfrak{F}_{\bb{p}}(\bb{r}_{12},\bb{r}_{31}) V(r_{12}) V(r_{31}) V(r_{12} \! + \! r_{31}) \biggr] \Bigl( \hat{a}_{\bb{p}}^{\dagger} \hat{a}_{\bb{p}} \! + \! \hat{a}_{-\bb{p}}^{\dagger} \hat{a}_{-\bb{p}} \Bigr) + \\
& + \frac{n_0^2}{3} \! \int \! \! \mathrm{d}\bb{r}_{12} \, \mathrm{d}\bb{r}_{31} \, \mathfrak{F}_{\bb{p}}(\bb{r}_{12},\bb{r}_{31})  V(r_{12}) V(r_{31}) V(r_{12} \! + \! r_{31})  \Bigl( \hat{a}_{\bb{p}}^{\dagger} \hat{a}_{-\bb{p}}^{\dagger} \! + \! \hat{a}_{\bb{p}} \hat{a}_{-\bb{p}} \Bigr)  \biggr\} \mbox{,}
\end{aligned}
\end{aligned}
\end{equation}
which is diagonalizable by the standard procedure. 
Then the excitation spectrum is:
\begin{equation}
\epsilon_p =  \sqrt{ \bigl(\epsilon_p^0\bigr)^2 + \frac{2}{3} n_0^2 \Bigl( \int \!  \mathrm{d}\bb{r}_{12} \, \mathrm{d}\bb{r}_{31} \, \mathfrak{F}_{\bb{p}}(\bb{r}_{12},\bb{r}_{31}) V(r_{12}) V(r_{31}) V(r_{12} + r_{31}) \Bigr) \, \epsilon_p^0 \biggr] }, 
\end{equation}
and we obtain
\begin{equation}
\epsilon_p = \sqrt{ \bigl(\epsilon_p^0\bigr)^2 + \frac{2}{3} n_0^2 \Bigl( \int \!  \mathrm{d}\bb{r}_{12} \, \mathrm{d}\bb{r}_{31} \Bigl(  e^{-\frac{i}{\hbar}\bb{p} \cdot \bb{r}_{12}} + e^{-\frac{i}{\hbar}\bb{p} \cdot \bb{r}_{31}} + e^{-\frac{i}{\hbar}\bb{p} \cdot (\bb{r}_{12} + \bb{r}_{31})} \Bigr) V(r_{12}) V(r_{31}) V(r_{12} + r_{31}) \Bigr) \, \epsilon_p^0 \biggr] \mbox{.} \label{eqn:spectrum_CV}
}
\end{equation}

To demonstrate the equivalence between Eqs. \eqref{eqn:spectrum_CV} 
and \eqref{eqn:spectrum_FT} we have to show that their prefactors 
multiplying $\epsilon_p^0$ are equal. 
We start considering the factor of Eq. \eqref{eqn:spectrum_CV}, 
which we denote by $f_{\bb{p}}$:  
\begin{equation}
f_{\bb{p}} \equiv \frac{2}{3} n_0^2 \! \int \!  \mathrm{d}\bb{r}_{12} \, \mathrm{d}\bb{r}_{31} \Bigl(  e^{-\frac{i}{\hbar}\bb{p} \cdot \bb{r}_{12}} + e^{-\frac{i}{\hbar}\bb{p} \cdot \bb{r}_{31}} + e^{-\frac{i}{\hbar}\bb{p} \cdot (\bb{r}_{12} + \bb{r}_{31})} \Bigr) V(r_{12}) V(r_{31}) V(r_{12} + r_{31}) \mbox{.}
\label{eqn:appendix_1}
\end{equation}
Since
\begin{equation}
V(r_{12})V(r_{31})V(r_{12} \! + \! r_{31}) = \frac{1}{\Omega^3} \! \! \sum_{\substack{\bb{p}_{12} \\ \bb{p}_{31} \\ \bb{p}_{12+31}}} \! \! V_{\bb{p}_{12}} \, V_{\bb{p}_{31}} \, V_{\bb{p}_{12+31}} \, e^{\frac{i}{\hbar} \bb{p}_{12} \cdot \bb{r}_{12}} \, e^{\frac{i}{\hbar}  \bb{p}_{31} \cdot \bb{r}_{31}} e^{\frac{i}{\hbar}  \bb{p}_{12 + 31} \cdot ( \bb{r}_{12} + \bb{r}_{31} )}  \mbox{,}
\end{equation} 
inserting this expression in \eqref{eqn:appendix_1}, 
it is readily seen that $f_{\bb{p}}$ is given by
\begin{equation}
f_{\bb{p}} = \frac{2}{3} \frac{n_0^2}{\Omega} \! \! \sum_{\substack{\bb{p}_{12} \\ \bb{p}_{31} \\ \bb{p}_{12+31}}} \! \! V_{\bb{p}_{12}} \, V_{\bb{p}_{31}} \, V_{\bb{p}_{12+31}} \biggl\{  \delta_{\bb{p},\bb{p}_{12}+\bb{p}_{12+31}} \delta_{\bb{p}_{31},-\bb{p}_{12+31}} + \delta_{\bb{p}_{12},-\bb{p}_{12+31}} \delta_{\bb{p},\bb{p}_{31}+\bb{p}_{12+31}} + \delta_{\bb{p},\bb{p}_{12}+\bb{p}_{12+31}} \delta_{\bb{p},\bb{p}_{31}+\bb{p}_{12+31}}    \biggr\} \mbox{,}
\end{equation}
which is equal to
\begin{gather}
f_{\bb{p}} = \frac{2}{3} \frac{n_0^2}{\Omega} \sum_{\bb{p}_{12}} \biggl\{ V_{\bb{p}_{12}} \, V_{\bb{p}_{12} - \bb{p}} \, V_{-\bb{p}_{12} + \bb{p}} + V_{\bb{p}_{12}} \, V_{\bb{p} + \bb{p}_{12}} \, V_{-\bb{p}_{12}} + V_{\bb{p}_{12}}^2 \, V_{-\bb{p}_{12} + \bb{p}}  \biggr\} \mbox{.}
\end{gather}
Using in the previous expression the properties 
$\sum_{\bb{p}} V_{\bb{p}} = \sum_{\bb{p}} V_{- \bb{p}}$ and 
\begin{equation}
\sum_{\bb{p}_{12}} V_{\bb{p}_{12}}^2 \, V_{\bb{p}_{12} + \bb{p}} = \sum_{\bb{p}_{12}} V_{\bb{p}_{12}}^2 \, V_{-\bb{p}_{12} + \bb{p}} \mbox{,}
\end{equation}
we retrieve the prefactor of the $\epsilon_p^0$ term 
in Eq. \eqref{eqn:spectrum_FT}.

\section{Thomas-Fermi approximation for the cubic-quintic 
Gross-Pitaevskii in an isotropic parabolic trap} 
\label{sec:TF_is}

In this Appendix we consider the time-independent cubic-quintic 
Gross-Pitaevskii equation
\begin{equation}
-\frac{\hbar^2}{2m} \nabla^2 \psi + V \psi + g_2 |\psi|^2 \psi + 
g_3 |\psi|^4 \psi = \mu \psi \mbox{,}
\label{cub_qui}
\end{equation}
where the parabolic trap $V(\bb{r})$ is assumed isotropic:
\begin{equation}
V(\bb{r})=\frac{1}{2} m \omega^2 (x^2+y^2+z^2) \mbox{.}
\end{equation}
To make contact with the notation used in the main text, 
it is $U_2=g_2$ and $g_3=U_3/2$. 

In the Thomas-Fermi approximation \cite{pethick2002,pitaevskii2016} 
it is found with $g_3>0$ that
\begin{equation}
|\psi(\vec{r})|^2=\frac{\sqrt{g_2^2+4g_3 [\mu-V(\vec{r})]}-g_2}{2g_3} \mbox{.}
\end{equation}

Imposing the normalization condition 
$\int d\vec{r} \, |\psi|^2=N$ and defining the Thomas-Fermi radius $R$ such 
that $V(R)=\mu$, one gets
\begin{equation}
N=A {\cal R}^4 F\left( \frac{B}{{\cal R}^2}\right) - C {\cal R}^3 \mbox{,}
\label{TF}
\end{equation}
where ${\cal R}$ is measured in units of the harmonic oscillator length 
$a=\sqrt{\hbar/m \omega}$ (with ${\cal R} \equiv R/a$) and the dimensionless 
quantities $A$, $B$ and $C$ in Eq. \eqref{TF} are given by $A= \mathsmaller{2\pi} 
\sqrt{\frac{2m \omega^2}{g_3}} \mathsmaller{a^4}$, $B=\frac{g_2^2}{2g_3\hbar \omega}$ and 
$C=\frac{2\pi g_2a^3}{3g_3}$. In \eqref{TF} the function $F(x)$ is defined (with $x>0$) 
as 
\begin{equation}
F(x)=\int_0^1 \theta^2 \sqrt{1+x-\theta^2} d\theta = 
\frac{1}{8} \left[ \left(1-x\right) \sqrt{x} + \left( 1+x\right)^2 
\mathrm{arccsc} 
\left( \sqrt{1+x} \right) 
\right] \mbox{.} 
\label{func}
\end{equation}

The expectation value of 
$\langle r^2 \rangle=\frac{\int d\vec{r} r^2 |\psi|^2}{N}$ 
is given by
\begin{equation}
\frac{\langle r^2 \rangle}{R^2}=\frac{A {\cal R}^4 G\left( \frac{B}{{\cal R}^2}\right) - \frac{3}{5}C {\cal R}^3}{N} \mbox{,}
\label{val_med_r2}
\end{equation}
where $R$ is defined by Eq. \eqref{TF} and the function $G$ is given by
\begin{equation}
G(x)=\int_0^1 \theta^4 \sqrt{1+x-\theta^2} d\theta = 
\frac{1}{48} \left[ \left(1-3x\right)\left(3+x\right) \sqrt{x} + 
3 \left( 1+x\right)^3  
\mathrm{arccsc} \left( \sqrt{1+x} \right) \right] \mbox{.} 
\label{func2}
\end{equation}
Since $\langle r^2 \rangle=\langle x^2 \rangle+
\langle y^2 \rangle+\langle z^2 \rangle$, we get 
$\langle x^2 \rangle=\langle y^2 \rangle=\langle z^2 \rangle=
\langle r^2 \rangle/3$. To adapt the notation to that of the main text, 
we set $\sigma_\alpha \equiv \sqrt{\langle \alpha^2 \rangle}$ and 
$\ell_\alpha \equiv 2\sigma_{\alpha}$, with $\alpha=x,y,z$, so that in the isotropic 
case considered in this Appendix the product 
$\ell_x \ell_y \ell_z$, 
which we denote by $\Omega_{TF}$, reads
\begin{equation}
\Omega_{TF}= \frac{8}{3\sqrt{3}} \, \langle r^2 \rangle^{3/2} \mbox{,} 
\label{omega_TF}
\end{equation}
where $\langle r^2 \rangle$ is given by Eq. \eqref{val_med_r2}.

The previous formulas simplify for $g_2=0$, when only the $3$-body 
interaction term [i.e., only 
the quintic term in the Gross-Pitaevskii equation \eqref{cub_qui}] is present. 
One gets
\begin{equation}
\frac{R^4}{a^4}=\frac{4 N \sqrt{2 {\cal G}_3}}{\pi^2} \mbox{,} 
\label{R_qui}
\end{equation}
where we introduced the dimensionless parameter ${\cal G}_3 \equiv 
\frac{g_3}{\hbar \omega a^6}$ (remember that $[g_3]=[E] \, \cdot \, [L]^6$). The expectation value 
of $\langle r^2 \rangle$ is simply given by 
\begin{equation}
\langle r^2 \rangle=\frac{R^2}{2} \mbox{.}
\label{val_med_r2_qui}
\end{equation}
Eqs. \eqref{R_qui} and \eqref{val_med_r2_qui} have to be compared 
with the usual results for the cubic Gross-Pitaevskii (with $g_3=0$), 
where one has respectively 
\begin{equation}
\frac{R^5}{a^5}=\frac{15 N {\cal G}_2}{4 \pi}  \mbox{,}
\label{R_cub}
\end{equation}
(where ${\cal G}_2 \equiv 
\frac{g_2}{\hbar \omega a^3}$) and 
\begin{equation}
\langle r^2 \rangle=\frac{3 R^2}{7} \mbox{.}
\label{val_med_r2_qui_app}
\end{equation}


\end{document}